%% file: ms.tex
\documentclass{emulateapj}
\usepackage{lscape}
\usepackage{natbib}

\slugcomment{Accepted for Publication in ApJ}
\shorttitle{UV/Optical Detections of Candidate Tidal Disruption Events}
\shortauthors{Gezari et al.}

\newcommand {\apgt} {\ {\raise-.5ex\hbox{$\buildrel>\over\sim$}}\ }
\newcommand {\aplt} {\ {\raise-.5ex\hbox{$\buildrel<\over\sim$}}\ }

\newcommand{\mbh}{M_{\rm BH}}%
\newcommand{\msun}{M_{\sun}}%
\newcommand{\etal}{et al.~}%
\begin{document}

\title{UV/Optical Detections of Candidate Tidal Disruption Events \\ by \textsl{GALEX} and CFHTLS\altaffilmark{1}}

\author{S.~Gezari\altaffilmark{2,3}, S.~Basa\altaffilmark{3}, D.~C.~Martin\altaffilmark{2}, G.~Bazin\altaffilmark{4}, K.~Forster\altaffilmark{2}, B.~Milliard\altaffilmark{3}, J.~P.~Halpern\altaffilmark{5}, P.~G.~Friedman\altaffilmark{2}, P.~Morrissey\altaffilmark{2}, S.~G.~Neff\altaffilmark{6}, D.~Schiminovich\altaffilmark{5}, M.~Seibert\altaffilmark{7}, T.~Small\altaffilmark{2}, and T.~K.~Wyder\altaffilmark{2}}

\altaffiltext{1}{Some of the data presented herein were obtained at the W.~M.~Keck Observatory, which is operated as a scientific partnership among the California Institute of Technology, the University of California, and the National Aeronautics and Space Administration.  The Observatory was made possible by the generous financial suppeort of the W.~M.~Keck Foundation.}

\altaffiltext{2}{California Institute of Technology, 
        MC 405-47,
        1200 East California Boulevard,
        Pasadena, CA  91125 \email{suvi@srl.caltech.edu}}

\altaffiltext{3}{Laboratoire d'Astrophysique de Marseille, 
        13376 Marseille Cedex 12, France}

\altaffiltext{4}{DSM/DAPNIA, CEA/Saclay, 91191 Gif-sur-Yvette Cedex, France}

\altaffiltext{5}{Department of Astronomy, 
    Columbia University, 
        New York, NY  10027}

\altaffiltext{6}{Laboratory for Astronomy and Solar Physics, 
     NASA Goddard Space Flight Center, 
      Greenbelt, MD  20771}

\altaffiltext{7}{Observatories of the Carnegie Institute of Washington, 
813 Santa Barbara St., Pasadena, CA  90095}

\begin{abstract}
We present two luminous UV/optical flares from the nuclei of apparently inactive early-type galaxies at $z=0.37$ and $0.33$ that have the radiative properties of a flare from the tidal disruption of a star.   In this paper we report the second candidate tidal disruption event discovery in the UV by the \textsl{GALEX} Deep Imaging Survey and present simultaneous optical light curves from the CFHTLS Deep Imaging Survey for both UV flares.  The first few months of the UV/optical light curves are well fitted with the canonical $t^{-5/3}$ power-law decay predicted for emission from the fallback of debris from a tidally disrupted star.   \textsl{Chandra} ACIS X-ray observations during the flares detect soft X-ray sources with $T_{bb}= (2-5) \times 10^{5}$ K or $\Gamma > 3$ and place limits on hard X-ray emission from an underlying AGN down to $L_{X}(2-10$ keV) $\aplt 10^{41}$ ergs s$^{-1}$.  Blackbody fits to the UV/optical spectral energy distributions of the flares indicate peak flare luminosities of $\apgt 10^{44}-10^{45}$ ergs s$^{-1}$.  The temperature, luminosity, and light curves of both flares are in excellent agreement with emission from a tidally disrupted main-sequence star onto a central black hole of several times $10^{7} \msun$.   The observed detection rate of our search over $\sim$ 2.9 deg$^{2}$ of \textsl{GALEX} Deep Imaging Survey data spanning from 2003 to 2007 is consistent with tidal disruption rates calculated from dynamical models, and we use these models to make predictions for the detection rates of the next generation of optical synoptic surveys.  
\end{abstract}

\keywords{galaxies: nuclei --- Ultraviolet: galaxies -- X-rays: galaxies}

\section{Introduction \label{intro}}

The tidal disruption of a star by a supermassive black hole (SMBH) has been proposed to be a unique probe for dormant black holes lurking in the nuclei of normal galaxies \citep{fr76,lid79}.  When a star passes close enough to a central black hole to be torn apart by tidal forces, the stellar debris falls back onto the black hole and produces a luminous accretion flare \citep{rees88}.  The detection of a tidal disruption flare is the only direct signpost for dormant black holes in the nuclei of galaxies for which the dynamical signature of the SMBH cannot be measured, i.e. when the sphere of influence of the black hole is unresolved ($R_{inf}=G\mbh/\sigma_{\star}^2$).  

Tidal disruption events are a potential new tool for measuring the SMBH mass function, independent of the scaling relations between central black hole mass and global host galaxy properties established for nearby galaxies with spatially resolved kinematics \citep{geb00, ferr00, gra01, marc03}, and the scaling relations measured for nearby active galaxies with black hole masses from reverberation mapping \citep{wand99, kasp00, vest02}. Tracing the coevolution of the mass of black holes and their host galaxies as a function of redshift is critical for constraining mechanisms for black hole growth and the formation of the host galaxy bulge, as well as the interplay between them \citep{kauff00, catt05, crot06}.

\subsection{Theory \label{sec:theory}}

The rate at which stars pass within the tidal disruption radius of a central
SMBH, $R_{T} =  R_{\star}(\eta^{2} \mbh/M_{\star})^{1/3}$, where $\eta$ is on the order of unity, depends on the flux of stars into loss cone orbits with $R_{p} < R_{T}$, where $R_{p}$ is the distance of closest approach of the star's orbit.  Stellar dynamical models of the nuclei of galaxies \citep{mt99,wm04} predict that this will occur
once every 10$^3$--10$^{5}$ yr in a galaxy, depending on the nuclear stellar density profile and the mass of the central black hole.  At these rates, stellar disruption events should contribute to the faint end of the X-ray luminosity function of active galaxies, and may be an important mechanism for black hole growth in black holes with $\mbh < 2 \times 10^{6} \msun$ \citep{milo06}, as well as a way to regulate the tight correlation between $\mbh$ and the velocity dispersion host galaxy bulge \citep{zhao02, merr04}.  For higher mass black holes, however, there is a critical black hole mass above which $R_{T}$ is smaller than the Schwarzschild radius ($R_{s}$),
and a star is swallowed whole without disruption \citep{hills75}, $M_{crit} = 1.15 \times 10^{8} (r^3/m )^{1/2} \msun$, where $r_{\star}=R_{\star}/R_{\odot}$ and $m_{\star}=M_{\star}/M_{\odot}$.  

After disruption, the stellar debris is distributed in total energy from $-\Delta\epsilon<E<+\Delta\epsilon$, where $\Delta\epsilon=G\mbh R_{\star}/R_{p}^2$ is the change in the black hole potential across the radius of the star \citep{lacy82}.  If $\Delta\epsilon \gg \epsilon_{b} = GM_{\star}/R_{\star}$, where $\epsilon_{b}$ is the mean orbital binding energy of the stellar debris, half of the mass of the star remains bound to the black hole, and half is ejected from the system.  Numerical simulations show that the fraction of debris that is eventually accreted can be as little as 0.1M$_{\star}$ due to the strong compression of the gas stream during the second passage through pericenter which can impart a fraction of the gas with an escape velocity \citep{ayal00}.  The most tightly bound gas has a total energy $E = G\mbh/r_{a}$ = $-\Delta\epsilon$, where $r_{a}$ is the apocentric radius, and thus $r_{a} = k^{-1} R_{\star}(\mbh/M_{\star})^{2/3}$, where $k=1$ if the star has no spin on disruption and $k=3$ if the star is spun-up to near break-up on disruption.  Numerical simulations predict that the star should be spun up via tidal interaction to a significant fraction of the break-up spin angular velocity \citep{alex01}.  The start of the flare, $t_{0}$, occurs when the most tightly bound gas returns to pericenter in a time after the disruption, $t_{D}$, where $t_{0}-t_{D} = \pi (r_{a}^3/(2G\mbh))^{1/2}$, and thus

\begin{equation}
$$\[t_{0}-t_{D} = 0.11 k^{-3/2} (R_{p}/R_{T})^3 
r_{\star}^{3/2} m_{\star}^{-1} M_{6}^{1/2} {\rm yr}\]$$ 
\label{eq:tmin}
\end{equation}

\noindent where $M_{6} = \mbh/10^{6} \msun$ \citep{li02}.  This time delay can be used as a direct measure of the central black hole mass, since $\mbh \propto k^{3}(t_{0}-t_{D})^2$.  

The accretion rate from the fallback of the most tightly bound debris is predicted to result in a flare which peaks close to the Eddington luminosity, with a blackbody spectrum with $T_{eff} \approx
(L_{Edd}/4\pi\sigma R_{T}^{2})^{1/4} = 2.5 \times 10^{5} M_{6}^{1/12}r_{\star}^{-1/2}$ K, which peaks in the extreme-UV \citep{ulmer99}.  The rate at which the remaining debris falls-back to pericenter is $dM/dt = (dM/d\epsilon)(d\epsilon/dr_{a})(dr_{a}/dt)$  \citep{phi89}.  If the mass is equally distributed in binding energy, which is shown to be a reasonable approximation from numerical simulations \citep{evans89, ayal00}, i.e. $dM/d\epsilon$ is constant, then $dM/dt \propto (G\mbh)^{2/3} t^{-5/3}$.  During the fallback phase, the rate of gas supplied to the black hole follows this ($t-t_{D})^{-5/3}$ power-law decay, and determines the luminosity of the flare over the following months and years.  

Thermal emission is predicted to be produced during fallback from stream-stream
collisions of the disruption debris.  Relativistic precession of the debris orbits causes the ingoing and outgoing gas streams to collide \citep{rees88, can90, kim99}, resulting in strong shocks that quickly circularize the debris.  The timescale for the debris to shock and circularize and form an accretion torus should be on the order of a few times the minimum period of the debris, given in equation 1 \citep{can90}.  At later times, the stellar debris is expected to spread out and form a thin disk that accretes slowly via viscous processes, with a mass accretion rate with a shallower $(t-t_{D})^{-n}$ power-law decay that approaches $n=1.2$ \citep{can90}.  
It has also been proposed that an extended envelope will form around the disk from the expelled debris and reprocess the 
radiation with an effective temperature of $T_{eff} \approx 1.3 \times 10^{4} (\mbh/10^{6} \msun)^{1/4}$ K \citep{ulm98}.

\subsection{Previous Candidates}

Although the theoretical work described above makes predictions for the rate, luminosity, temperature, and decay of tidal disruption events, there is limited observational evidence to test these predictions.  The most convincing cases for a stellar disruption event occur from host galaxies with no evidence of an active galactic nucleus (AGN) for which an upward fluctuation in the accretion rate could also explain a luminous UV/X-ray flare.  A UV flare with $L \sim 10^{39}$ ergs s$^{-1}$ from the nucleus of the elliptical galaxy NGC 4552 was proposed to be the result of the accretion of a tidally stripped stellar atmosphere \citep{ren95} by the mini AGN at its center \citep{capp99, xu05}.  A more exotic candidate for a tidal disruption event was discovered in the central region of a rich cluster:  a transient optical source whose colors and light curve were best explained as either a peculiar underluminous Type Ia supernova (SN Ia) at the redshift of the cluster ($z\approx0.25$), or a lensed tidal disruption event at $z \approx 3.3$ \citep{stern04}.

The \textsl{ROSAT} All Sky Survey \citep{voges99} conducted in 1990--1991
sampled hundreds of thousands of galaxies in the soft X-ray band, and detected luminous ($10^{42}-10^{44}$ ergs s$^{-1}$),
soft [$T_{\rm bb} = (6 - 12) \times 10^{5}$ K and $\Gamma = 3-5$] X-ray flares 
from several galaxies with no previous evidence
for AGN activity \citep{bade96, kombade99, komgre99, grupe99, gre00} which were explained as tidal disruption events (see review by Komossa \etal 2002), and with a flare rate of $\approx 1 \times 10^{-5}$ yr$^{-1}$ per galaxy \citep{donley02}, which is in agreement with
the theoretical stellar disruption rate.  
A decade later, follow-up \textsl{Chandra} and \textsl{XMM}-Newton observations of three of the galaxies demonstrated that they had
faded by factors of $240 - 6000$.  
Although this dramatic fading is consistent 
with the ($t-t_{D})^{-5/3}$ decay of a tidal disruption flare \citep{kom04, hgk04, li02},
the follow-up observations were obtained too long after the peak of the flares (5--10 yr) to measure the shape
of the light curves.  Follow-up \textsl{Hubble Space Telescope (HST)} Space Telescope Imaging
Spectrograph (STIS) narrow-slit
spectroscopy confirmed two of the galaxies as inactive, qualifying them as the 
most convincing hosts of a tidal disruption event \citep{gez03}.    

The soft X-ray outburst from NGC 5905, which has
the best sampled light curve of the \textsl{ROSAT} candidates, was fitted with a $t^{-5/3}$ power-law decay (Komossa \& Bade 1999),  and the energetics of the flare was successfully modeled as emission from the tidal disruption of a brown dwarf or giant planet \citep{li02}.  Alternative scenarios for the source of the flare in NGC 5905, such as an SN, gamma-ray burst, lensing event, or variable absorbed Seyfert nucleus, were ruled out by Komossa \& Bade (1999).  NGC 5905 was subsequently found to have a persistent low-luminosity Seyfert nucleus in its STIS spectrum (Gezari \etal 2003).  However, the amplitude of its X-ray variability ($> 1000$) is better attributed to a tidal disruption event than extreme variability of its low-luminosity active nucleus.  

It should be noted that the presence of a low-luminosity AGN and the occurrence of a tidal disruption event are not mutually exclusive.  It is possible that the presence of an accretion disk may even enhance the tidal disruption rate due to interactions between a star and disk that decrease its angular momentum and bring it into a loss-cone orbit \citep{sy91, arm96, donley02}. Optical spectroscopic surveys reveal that $20-40\%$ of nearby galaxies host a Seyfert nucleus \citep{ho97, mill03}.  Low levels of AGN activity may be a ubiquitous property of galaxies with a central SMBH.  However, the presence of an active nucleus does make the interpretation
of a UV/X-ray flare subject to careful analysis.

Two new candidates from the XMM-Newton Slew Survey (XMMSL1) were reported by \citet{esq06} to be optically non-active galaxies that demonstrated a large-amplitude (80--90) increase in soft (0.1-2 keV) X-ray luminosity compared to their previous \textsl{ROSAT} PSPC All-Sky Survey upper-limits, with XMMSL1 luminosities of $10^{41}-10^{43}$ ergs s$^{-1}$.  

With the exception of NGC 5905, all of the X-ray tidal disruption event candidates described above were simply ``off''-``on'' detections, with no detailed light-curve information.  We initiated a program to take advantage of the UV sensitivity,
large volume, and temporal sampling of the \textsl{Galaxy Evolution Explorer (GALEX)} Deep Imaging Survey (DIS) 
to search for stellar disruptions in the nuclei 
of galaxies over a large range of redshifts and attempt to measure the detailed properties of a larger sample of tidal disruption events.  

In \citet{gez06} we reported our first detection of a candidate tidal disruption flare by \textsl{GALEX}: a UV flare from an inactive early-type galaxy at $z=0.3698$ with a spectral energy distribution (SED) from the optical through the X-rays fitted by a blackbody temperature  of $(1-5) \times 10^{5}$ K, a blackbody radius of $\approx 10^{13}$ cm, a bolometric luminosity of $10^{43}-10^{45}$ ergs s$^{-1}$, and with a UV light curve well-fitted by a ($t-t_{D})^{-5/3}$ power-law, in excellent agreement with theoretical predictions for emission from the tidal disruption of a star.  In this paper we strengthen this interpretation even further by extending the UV light curve of the flare with another epoch of \textsl{GALEX} observations and by presenting its simultaneous optical light curve from the CFHTLS Deep Imaging Survey.  We also report a new candidate tidal disruption flare detection from an inactive early-type galaxy at $z=0.326$ by \textsl{GALEX} and present its simultaneous optical light curve from the CFHTLS.  The broadband properties of the two flares, from the optical through the X-rays, are in excellent agreement with the predictions of basic tidal disruption theory and are both the most detailed observations of candidate stellar disruption events to date.

The paper is organized as follows.  In \S2 we describe our techniques for selecting tidal disruption flare candidates, describe our sample of sources classified as quasars, and report the detection in the far-ultraviolet (FUV) and near-ultraviolet (NUV) of an SN IIn at $z=0.189 $ discovered by the CFHT Supernova Legacy Survey (SNLS); in \S3 we describe our tidal disruption flare candidates in the UV and optical; in \S4 we interpret the properties of the flares in the context of tidal disruption theory; in \S5 we use the observed properties of the flares to measure the efficiency of our search and calculate the expected tidal disruption flare detection rate from theoretical models; and in \S6 we predict the detection rate of tidal disruption events by the next generation of optical synoptic surveys.  Throughout this paper, calculations are made using 
\textsl {Wilkinson Microwave Anisotropy Probe (WMAP}) cosmological parameters \citep{ben03}: 
$H_{0}$ = 71 km s$^{-1}$ Mpc$^{-1}$, $\Omega_{M} = 0.27$, and $\Omega_{\Lambda} = 0.73$.  Rest-frame magnitudes are estimated using the IDL routine KCORRECT version 4.1.4 \citep{blan03}, and when necessary, CFHTLS magnitudes are converted to the SDSS system using the offset corrections measured by the SNLS \citep{reg07}.

\section{Selection of Candidates \label{sec:select}}
Our aim is to find UV flares from the nuclei of normal, quiescent galaxies, for which the flare is a one-time accretion event for an otherwise dormant or undetectable central black hole.  Such flares are best explained as the result of the disruption and accretion of a star.
We select our tidal disruption flare candidates from variable UV sources in the \textsl{GALEX} Deep Imaging Survey (DIS) that have optical matches in the CFHTLS Deep Imaging Survey with the optical colors and morphology of a galaxy.  We use available X-ray catalog data to rule out galaxy hosts with hard X-ray emission from an AGN and obtain optical spectroscopy to eliminate galaxies with emission lines that manifest Seyfert activity in their nuclei.  We also match the variable UV sources to the list of variable optical sources from the CFHT SNLS to check for anti-coincidence with SN triggers. 

The DIS covers 80 deg$^{2}$ of sky in the FUV ($\lambda_{eff}=1539 $\AA) and NUV ($\lambda_{eff}=2316 $\AA) with a total exposure time of 30--150 ks, 
which is accumulated in visits during $\sim 1.5$ ks eclipses, when the satellite's 98.6 minute 
orbit is in the shadow of the Earth \citep{martin05, mor05, mor07}. Due to target visibility
and mission planning constraints, some 1.2 deg$^{2}$ DIS fields are observed over a baseline of $2-4$ yr to complete the total exposure time.  This large range in cadence of the observations
allows us to probe variability on timescales from hours to years.  Here we present four DIS fields that overlap with the optical Canada-France-Hawaii Telescope Legacy Survey (CFHTLS) Deep Imaging Survey\footnote{Described in http://www.cfht.hawaii.edu/Science/CFHLS.}: XMM/LSS (CFHTLS D1), COSMOS (CFHTLS D2), GROTH (CFHTLS D3), and CFHTLS D4.  

In order to optimize our sensitivity to UV flares that decay on the timescale of months to years, we co-added the visits of each field into yearly epochs using the \textsl{GALEX} pipeline.  Table \ref{tab:texp} gives the coordinates and exposure times of the yearly co-added images in the FUV and NUV for each field from 2003 through 2007.  To avoid the edges of the images that have degraded astrometry and photometry, as well as edge reflections, we only use sources with $r \le 0\fdg55$ from the center of the field.   We search for variable sources in the FUV co-added images, instead of the NUV, (1) in order to avoid blending between sources, which is a larger issue in the NUV; and (2) for better contrast between the variable source and the host galaxy, since galaxies are fainter in the FUV.  We use aperture magnitudes  ($r_{ap} = 6 \arcsec$) from the pipeline-generated SExtractor catalogs \citep{ber96} and match them to the closest source within a 2$\arcsec$ radius in the other co-added images, a radius that is less than half of the angular resolution in the FUV (FWHM= $4.5\arcsec$).  We only compare co-added images of a field with $t_{exp} > 5$ ks, and measure the standard deviation of sources that are detected in all the co-added images from the weighted mean magnitude, $\sigma(\langle m\rangle)$,
where $\langle m \rangle =\sum_{i=1}^N \frac{m_{i}}{\sigma_{i}^2}/\sum_{i=1}^N \frac{1}{\sigma_{i}^2}$, $N$ is the number of co-added images, $m_{i}$ are the matched magnitudes, and $\sigma_{i}$ are the Poisson errors,
 
\begin{equation}
$$\[\sigma_{i}= \frac{2.5}{f_{m}}\frac{1}{\ln{(10)}}\frac{\sqrt{(f_{m}+B_{sky}N_{pix})T_{exp}}}{T_{exp}},\]$$
\end{equation}

\noindent where $f_{m}$ is the aperture flux in counts s$^{-1}$, $B_{\rm sky}$ is the sky background at the position of the source in counts s$^{-1}$ pixel$^{-1}$ converted from the pipeline value for the sky background in counts arcsec$^{-2}$ (1 pixel = 1.5$^2$ arcsec$^2$), and $N_{pix}$ = 16$\pi$ for a 6$\arcsec$ radius aperture (1.5$\arcsec$ pixel$^{-1}$).  The measured standard deviation of matched magnitudes between co-added images is dominated by photometric errors.  For comparison, we also determine $\sigma_{phot}$ for matched sources, 

\begin{equation}
$$\[\sigma_{phot}(\langle m \rangle)=\sqrt{\frac{\sum_{i=1}^N \sigma_{i}^2}{N}},\]$$
\end{equation}

\noindent and the mean $\sigma_{phot}$ is measured as a function of magnitude in bins of magnitude with at least 100 points.  The curves are in good agreement with each other; however, to be conservative, we use the measured $\sigma$ in case there are systematic errors between co-added images not accounted for by $\sigma_{phot}$. 

We use a 5 $\sigma$ cutoff to select intrinsically variable sources.    The source counts start to fall off after the limiting magnitude of the shallowest coadd, so we do not use $\sigma$ measured for magnitudes fainter than $m_{lim}$.  We define the limiting aperture magnitude of an exposure by the 5$\sigma$ detection limit for a point source:

\begin{equation}
\label{eq:mlim}
$$\[m_{lim}= -2.5\times {\rm log}\left(5 \frac{\sqrt{B_{sky} N_{pix}T_{exp}}}{T_{exp}}\right) + {\rm zp},\]$$
\end{equation}

\noindent where $B_{sky} = 3 \times 10^{-4}$ cts/sec/pix is the typical sky background in the FUV, $B_{sky} = 3 \times 10^{-3}$ cts/sec/pix in the NUV, zp=18.82 in the FUV, and zp=20.08 in the NUV.  For this step in the analysis, we do not include an aperture correction for the fraction of total energy enclosed in the aperture.  

A large, extended galaxy can be shredded by SExtractor into multiple sources.  In this case, a source from a shredded galaxy may not have a match in another co-added image if the galaxy has not been ``shredded'' in the same way in the other coadd.  To avoid this problem, when a source in one co-added image ($C_{1}$) has no match in another co-added image ($C_{2}$), we place an aperture on $C_{2}$ to measure the aperture magnitude.  If the measured magnitude in $C_{2}$ is brighter than $m_{lim}$ of $C_{2}$ then we assume that the source in $C_{1}$ has been shredded from an extended galaxy, and that the lack of a source match in $C_{2}$ is not due to variability.

The 5$\sigma$ UV variable sources are then matched with the TERAPIX CFHTLS-T0003 Deep Survey photometric catalog in $u~(\lambda_{eff}=3743$ \AA), $g~(\lambda_{eff}=4872$ \AA), $r~(\lambda_{eff}=6282$ \AA), $i~(\lambda_{eff}=7776$ \AA), and $z~(\lambda_{eff}=11702$ \AA) and the optical colors and morphology of the matched sources are used to identify the hosts of the flares.  We use the closest match within a 3$\arcsec$ radius, which has been determined from other \textsl{GALEX} studies to be the optimum radius to match with optical catalogs of similar depth, with only 1\% of UV sources with multiple optical matches within that radius \citep{arg05}.  Tables \ref{xmm_tab1}--\ref{d4_tab1} list the FUV and NUV AB magnitudes in each co-added image of the UV variable sources that have optical matches, as well as the factor times $\sigma(m)$ that the source varies.  The magnitude errors given are 1$\sigma$ statistical errors, and do not take into account the 2\% and 5\% zeropoint plus flat fielding uncertainty in the NUV and FUV, respectively \citep{mor07}.  The aperture magnitudes have been corrected for the fraction of energy enclosed in a 6$\arcsec$ radius aperture, $m_{tot}=m_{ap}+2.5\log(f_{ap})$, where $f_{ap}$ is 0.95 and 0.86 for the FUV and NUV, respectively, measured from a composite point spread function (PSF) constructed in a DIS coadd.
Sources are flagged as flares (F) in Tables \ref{xmm_tab1}--\ref{d4_tab1} if they have a constant flux within the errors before the FUV flare, the flare fades monotonically thereafter, and it does not fade fainter than its flux before the flare.  If a monotonic FUV flare is undetected before the start of the flare, it is flagged in the tables as a transient flare (T).  

We designate optically resolved sources as those with a half-light radius ($r_{1/2}) > 0\farcs6$. One source (D3-15) is mis-classified as resolved by the optical catalog due to its bright magnitude ($m_{r} < 15$) which produces bright diffraction spikes, and this has been corrected in the table.  Optically unresolved sources with $g-r < 0.6$ and $u-g < 1$ are classified as quasars, unresolved sources with $u-g > 1.75$ or $g-r < 0.6$ and $u-g > 1$ are classified as stars, and all resolved sources are classified as galaxies.    X-ray catalog data from \textsl{XMM-Newton} for the XMM/LSS field \citep{chia05} and the COSMOS field \citep{man07} and from \textsl{Chandra} for the GROTH field \citep{nan05} are used to classify galaxies with a hard X-ray point-source detection as AGNs.  

\begin{figure}
\epsscale{.9}
\plotone{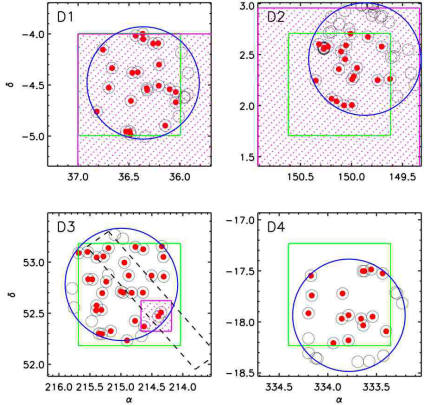}
\caption{{\textsl GALEX} variable sources ({\it open circles}) with CFHTLS optical matches ({\it red filled circles}).  The blue circle shows 0$\fdg$55 deg radius of {\textsl GALEX}, and the green square shows 1 deg square diameter of CFHTLS.   The area of overlap between the surveys is 0.806 deg$^{2}$ (D1), 0.565 deg$^{2}$ (D2),
0.802 deg$^{2}$ (D3), and 0.709 deg$^{2}$ (D4), for a total overlap of 2.882 deg$^{2}$.  The area of overlapping X-ray catalog data is shown shaded in magenta, and the perimeter of spectroscopic data from the DEEP2 Survey in the GROTH field is shown with a dashed black line.  \label{fig:over}}
\end{figure}

In addition, we match the sources to the list of spectroscopically confirmed quasars and active nuclei from \citet{ver06} and the COSMOS Survey \citep{trump07}.   We also match our sources with the list of quasars identified from \textsl{GALEX} UV grism spectra in the COSMOS and GROTH fields (T.~Small 2007, private communication).  Figure \ref{fig:over} shows the UV variable sources and their optical matches in four \textsl{GALEX} DIS fields and the overlap between \textsl{GALEX} and CFHTLS for each field, resulting in a total overlapping area of 2.882 deg$^{2}$.  Also shown is  the overlap of the available X-ray catalog data and the spectroscopic DEEP2 survey in the GROTH field.   

The CFHTLS fields are observed up to 5 times a month during the seasonal visibility of each field.  Real-time difference imaging is performed on the images to produce a list of optically variable sources from which the CFHT SNLS selects SN candidates for follow-up spectroscopy \citep{ast06,sull06}.  We match the publicly available variable optical source list (which includes optical variability associated with quasars and AGNs) with our UV variable sources.    Contamination by SN Ia and gamma-ray burst afterglows is unlikely since they are intrinsically faint in the UV \citep{pan03, brown05, mes97}.  However, SN II are UV bright at early times \citep{brown07, imm07} and are a potential contaminant, and so we check for anti-coincidence with the SNLS SN triggers.  

\begin{figure}
\epsscale{.9}
\plotone{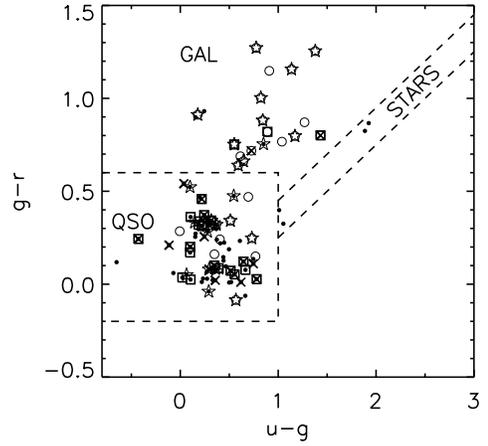}
\caption{ Optical color-color diagram of CFHTLS matches to {\textsl GALEX} UV variable sources.  Resolved optical hosts are plotted with open circles, and unresolved optical hosts are plotted with filled circles.  Sources detected by CFHT SNLS as variable sources in the optical are plotted with stars.  Sources with an X-ray catalog detection are plotted with crosses, and sources that are spectroscopically confirmed AGNs or quasars are plotted with squares.  Dashed lines delimit the regions of the diagram dominated by quasars, early-type galaxies, and main-sequence stars. \label{fig:all}}
\end{figure}

Figure \ref{fig:all} shows the optical color-color diagram of the UV variable source hosts and the regions of the diagram dominated by template CFHTLS colors for main-sequence stars, quasars, and early-type elliptical and spiral galaxies from the photometric redshift code Le PHARE.\footnote{Arnouts \& Ilbert, http://www.oamp.fr/people/arnouts/LE\_PHARE.html}  Our UV variability selection detects mostly quasar hosts, many of which are also detected as hard X-ray sources and as optically variable sources.   Four of the optically unresolved hosts have optical colors of a main-sequence star; however, three of these sources (D1-4, D2-6, D3-1) have a $>1\farcs0$ separation from the UV source, which is larger than the positional uncertainty of \textsl{GALEX} and may be a false match.  Our analysis does not detect any RR Lyrae stars, which are located below the main sequence with $u-g\sim$1.15 \citep{ses07}, or M-dwarf flare stars, located at the top of the main sequence with $u-g\sim$2.6 \citep{haw02}, since our deep co-added images average out their variability signal which occurs on much shorter timescales of days and minutes, respectively \citep{welsh05,welsh06}.  

Tables \ref{xmm_tab2}--\ref{d4_tab2} give the optical AB magnitudes, half-light radii, match separation of the flare hosts, and their classification described above.  Sources are flagged in the tables with an `o' if they are optically variable, with an `x' if they are detected in an X-ray catalog, with an `s' if they are confirmed AGNs or quasars from optical spectroscopy, and with a 'g' if they have a quasar spectrum in the UV.   The redshift of the host galaxy is also listed in the table for those with spectroscopy.  The UV variability of D2-21 is due to an artifact in the 2005 COSMOS coadd, and so it is labeled in Table \ref{cosmos_03_tab2} as an artifact (ART).  Due to errors in the centroid of the faint UV source D3-14, it was erroneously not matched with its counterpart in the co-added images after 2003 which had centroids that are offset by $2.8\arcsec$.  This source is mistakenly detected as a variable source and is labeled as an error (ERR) in Table \ref{groth_tab2}.

We define our tidal disruption event candidates, listed in Table \ref{tab_cand}, as monotonic or transient UV flares from optically resolved galaxy hosts that are not classified as an AGN (no hard X-ray detection and no existing AGN optical spectrum).
We follow up these galaxy hosts with optical spectroscopy, to search for signs of Seyfert activity in their optical spectra in the form of emission lines.  If a galaxy does not have a Seyfert nucleus, and thus no obvious signs of persistent accretion activity by its central black hole, the source of the UV flare is interpreted in the context of the tidal disruption of a star by a central black hole that is otherwise starved of gas and dormant.

\subsection{Quasars Selected from UV Variability and Optical Colors and Morphology}
\begin{figure}
\epsscale{.9}
\plotone{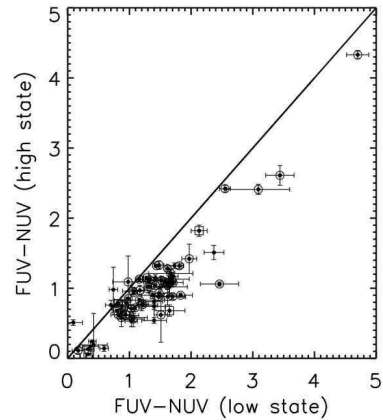}
\caption{FUV$-$NUV color of {\textsl GALEX} variable sources in their low state and high state of flux.  Sources classified as quasars are plotted with an open circle, and sources classified as AGNs are plotted with an open square.  All of the quasars and AGNs lie below the line of constant FUV$-$NUV color within the errors, indicating a bluer spectrum during high states of flux. \label{fig:3b}}
\end{figure}

As a biproduct of this study we have a list of 54 quasars classified from UV variability and optical colors and morphology alone.
Only 11\% of these quasars are detected as optically variable sources by the SNLS database.  This discrepancy should not be due to difficulties in detecting variable sources in the nuclei of galaxies with image subtraction, since these sources are by definition optically unresolved.  The SNLS optical variability search is optimized for detecting SNe, which vary on the timescale of weeks, and within the seasonal window of observability of the fields.  Longer timescale variability (months--years) may not be picked up by their selection techniques.
Our study demonstrates that UV variability in combination with optical colors and morphology is an efficient method for detecting quasars over a large field of view, without the expense of X-ray surveys and optical
spectroscopy.  
In Figure \ref{fig:3b} we plot the low-state and high-state FUV$-$NUV color for all the variable sources, with quasars and AGNs labeled.  Within the errors, all of the quasars and AGNs have a bluer FUV$-$NUV color during their high state in flux.  This is in agreement with the findings of optical variability studies \citep{giv99, geha03}.  This may also contribute to the better efficiency in detecting quasars with variability in the UV than in the optical.  

\subsection{Detection of a Supernova Type IIn}
Candidate D4-7 is an SN candidate that was discovered by the CFHT SNLS on 2003 August 21 (SNLS ID: 03D4ck), and was spectroscopically confirmed by Gemini on 2003 November 20 as an SN IIn at $z=0.189$ (Howell \etal 2005).  \textsl{GALEX} detects the SN in the 2003 co-added image of CFHTLS D4 constructed from visits from 2003 August 30 to 2003 September 21, with $FUV=23.43\pm0.08$ mag and $NUV=22.50\pm0.04$ mag.  This is the first UV detection of an SN IIn, and one of only 5 other published UV light curves of Type II SNe \citep[see][]{brown07}.  In a separate paper (L.~Dessart, in preparation) we present an analysis of the detailed UV light curve, along with the CFHT SNLS optical light curve and spectra.  

\subsection{Optical Spectroscopy of Candidates \label{sec:opt}}
\begin{figure}
\epsscale{.9}
\plotone{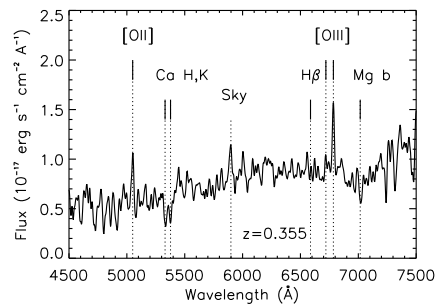}
\caption{MDM 2.4 m CCDS spectrum of the galaxy host of D3-3 that shows Seyfert-like narrow emission lines.  This galaxy
is also detected in an archival \textsl{Chandra} X-ray image with $L_{X} \sim 9.3 \times 10^{42}$ ergs s$^{-1}$. \label{fig:mdm}}
\end{figure}

\begin{figure}
\epsscale{.9}
\plotone{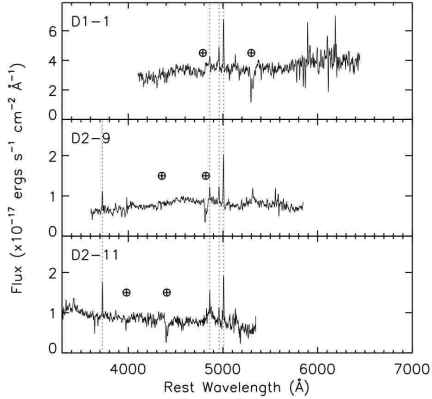}
\caption{Keck LRIS spectra of galaxy hosts of tidal disruption flare candidates shown plotted at their rest wavelength.  Hatch marks show the location of the strong O$_{2}$ telluric B- and A-band absorption features that are not corrected for in the spectra.  Dotted lines show the wavelengths of [O~II]$\lambda$3727, H$\beta$, and [O~III]$\lambda\lambda$4959,5007.  All three galaxies, D1-1, D2-9, and D2-11 have narrow H$\beta$/[O~III]$\lambda$5007 ratios and/or broad H$\beta$ emission lines that classify them as Seyfert galaxies.
\label{fig:lris1}}
\end{figure}

\begin{figure}
\epsscale{.9}
\plotone{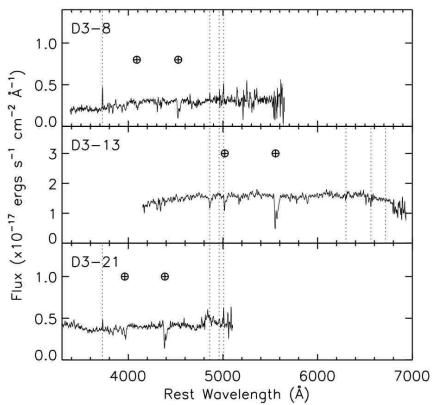}
\caption{Keck LRIS spectra of galaxy hosts of tidal disruption flare candidates shown plotted at their rest wavelength.  Hatch marks show the location of the strong O$_{2}$ telluric B- and A-band absorption features that are not corrected for in the spectra.  Dotted lines show the wavelengths of [O~II]$\lambda$3727, H$\beta$, and [O~III]$\lambda\lambda$4959,5007, [O~I]$\lambda$6300, H$\alpha$, and [S~II]$\lambda$6716.  D3-8 has a narrow H$\beta$/[O~III]$\lambda$5007 ratio that requires other narrow-line ratios to classify it.   D3-13 shows no sign of emission lines consistent with its classification as an inactive early-type galaxy by its AEGIS Keck DEIMOS spectrum. D3-21 has a broad H$\beta$ emission line that classifies it as a Seyfert 1 galaxy.\label{fig:lris2}}
\end{figure}

\begin{figure}
\epsscale{.9}
\plotone{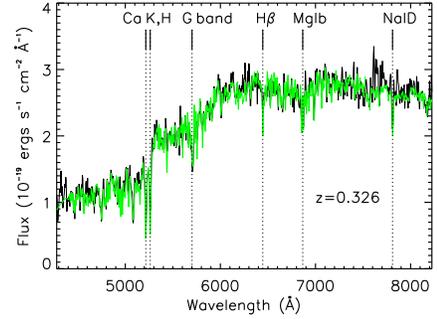}
\caption{ VLT FORS1 spectrum of the galaxy host of D1-9 obtained on 2006 September 26 smoothed by 4 pixels ($\sim$10 \AA).  The spectrum shows strong stellar absorption
lines typical of an early-type galaxy.  The best-fitting Bruzual-Charlot early-type galaxy template \citep{bruz03} with an old stellar population and no star formation is shown in green scaled to the continuum of the spectrum. \label{fig:spec}}
\end{figure}

Table \ref{tab:spec} gives the log of spectroscopic observations of the host galaxies of 8 out of 10 of our tidal disruption flare candidates, as well as their redshifts, ([O~III] $\lambda5007$)/(H$\beta$) narrow-line ratio, and [O~III] $\lambda5007$ luminosity.  We define a spectrum as a Seyfert 2 (Sy2) in the table if it has a diagnostic narrow-line ratio ([O~III] $\lambda5007$)/(H$\beta) > 3$ \citep{bald81, veill87, kew01} and as a Seyfert 1 (Sy1) if it shows broad H$\beta$ emission.   The spectra are shown in 
Figures \ref{fig:mdm}--\ref{fig:spec}.  D1-1, D2-9, and D3-3 have narrow-line ratios indicative of a Seyfert spectrum .  D3-3 is a Seyfert 2 galaxy at $z=0.355$ that is also detected in an archival \textsl{Chandra} ACIS image in 2002 April 10 indicating a hard X-ray luminosity of $L_{X}$(2-10 keV) $\sim9.3 \times 10^{42}$ ergs s$^{-1}$.  Three of the galaxies, D2-11, D3-8, and D3-21, have $2 < $([O~III] $\lambda5007$)/(H$\beta$)$ < 3$, which places them in a region of the diagnostic diagram that needs other line ratios to determine the class of the galaxy.  However, D2-9 and D3-21 have a broad H$\beta$ line, and so we classify them as Seyfert 1 galaxies.   D1-9 and D3-13 have spectra with only stellar absorption lines, and are classified as early-type galaxies.  This leaves D3-8 as the only galaxy without a classification from its optical spectrum. 

Table \ref{tab_cand} lists all of our candidates, the type of FUV flare (monotonic or transient), whether they were identified as optically variable in the SNLS database,  their photo-$z$ and best-fitting galaxy template from CFHTLS \citep{ilb06, brun07}, as well as the spectroscopic redshifts and classification for candidates with spectra. 
The photo-$z$ template fitting accurately classifies the early-type galaxies and emission-line galaxies, with the exception of D3-3, whose weak optical emission lines are not detected in the photo-$z$ fitting.      
D1-9 (Figure \ref{fig:spec}) and D3-13 (Figure \ref{fig:lris2}) show no emission lines in their spectra, and strong stellar absorption features and thus are classified as inactive, early-type galaxies.  We place upper limits on their [O~III]$\lambda 5007$ luminosities of $ < 5 \times 10^{38}$ and $ < 4 \times 10^{39}$ ergs s$^{-1}$, respectively.

\subsection{Optical Light Curves of Candidates}
\begin{figure}
\epsscale{.9}
\plotone{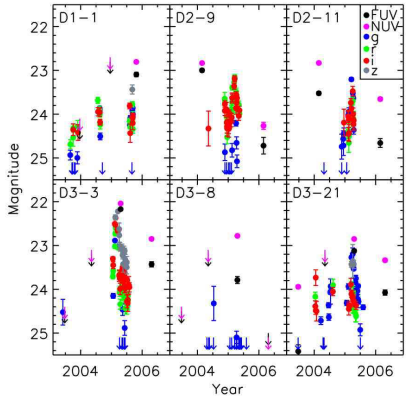}
\caption{UV/optical light curves of candidates with emission lines.  For non-transient UV flares, the low-state flux has been subtracted from the FUV and NUV fluxes.  The $g$, $r$, $i$, and $z$ data points are measured from difference imaging with CFHTLS.  For clarity, upper-limits (plotted with arrows) are shown only for the FUV, NUV, and $g$ band. \label{fig:sy_all}}
\end{figure}

We extract optical light curves for all of our candidate flares from the CFHTLS Deep Imaging Survey. The optical photometry is measured from difference imaging following the method of \citet{al98} implemented by \citet{leg03} in the $g$, $r$, $i$, and $z$ bands.  The light curves are calibrated to the Vega system and must be converted to AB magnitudes using the offsets measured for MegaPrime/MegaCam.\footnote{http://www.cfht.hawaii.edu/Instruments/Imaging/Megacam/generalinformation.html}  Upper limits are determined for a $5 \sigma$ point-source detection based on the average observing conditions using the MegaCam Direct Imaging Exposure Time Calculator (DIET).\footnote{http://www.cfht.hawaii.edu/Instruments/Imaging/Megacam/dietmegacam.html}  Figure \ref{fig:sy_all} shows the simultaneous UV/optical light curves extracted from the CFHTLS of the candidates with emission-line galaxy hosts.  For non-transient flares in the UV, the low-state flux was subtracted from the FUV and NUV fluxes.  

Only one of the candidates, D3-3, has a flare with the sharp rise and monotonic decay expected from a tidal disruption event.  The color of the flare is very red, with its peak optical flux in the $z$ band, and with a slope from the $g$ to the $z$ band well fitted with a power-law, $F_{\nu} \propto \nu^{\alpha}$, with $\alpha = -0.77 \pm 0.07$, which is consistent with the power-law optical continuum measured for AGNs $(-2.0 \le \alpha \le -0.43$; Zheng \etal 1997; Dietrich \etal 2002).  Its Seyfert emission lines [L([O~III] = $2.5 \times 10^{40}$ ergs s$^{-1}$], hard X-ray luminosity ($L_{X} \approx 10^{43}$ ergs s$^{-1}$), and power-law optical continuum are most consistent with the UV/optical flare being a result of variability of its AGN continuum.

\begin{figure}
\epsscale{.9}
\plotone{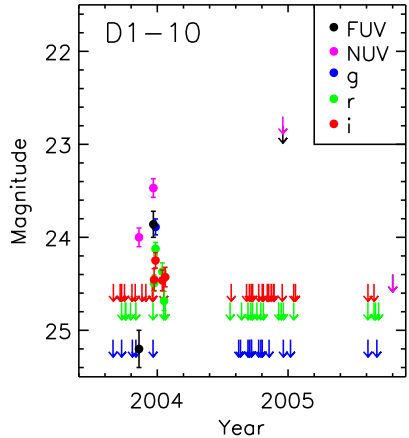}
\caption{UV/optical light curve of D1-10, a candidate without an optical spectrum.  The low-state flux has been subtracted from the FUV and NUV fluxes.  Arrows show upper limits.  The short duration of the flare and its offset from the nucleus of $0.6\arcsec$ are most likely associated with an SN II. \label{fig:d1_10}}
\end{figure}

\begin{figure}
\epsscale{.9}
\plotone{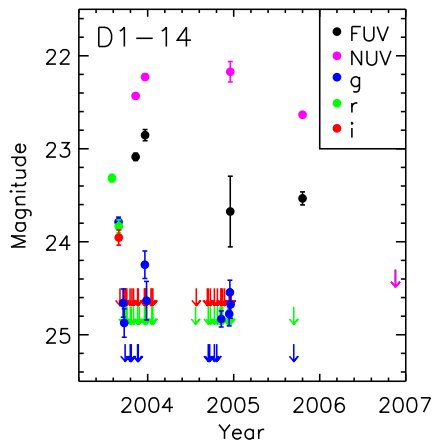}
\caption{UV/optical light curve of D1-14, a candidate without an optical spectrum.  The low-state flux has been subtracted from the FUV and $NUV$ fluxes.  Arrows show upper limits.  \label{fig:d1_14}}
\end{figure}

Our remaining candidates without optical spectroscopy, D1-10 and D1-14, have photo-$z$ galaxy templates that are consistent with an emission-line galaxy, although the emission lines could be from star formation or a Seyfert nucleus.  Figures \ref{fig:d1_10} and \ref{fig:d1_14} show the UV/optical light curve of the two candidates.  The short duration of the UV/optical flare in D1-10 of $\sim 70$ days, as well as the offset of the variable source from the nucleus of the galaxy measured by the SNLS database of $\sim 0.6\arcsec$, is compatible with an SN, and not a nuclear outburst.  The blue color of the flare, $NUV-r = -0.6 \pm 0.1$, and its location on the edge of a galaxy indicate that it is most likely an SN Type II, which are detected in the UV and are associated with star formation in the disks of galaxies. The high photo-$z$ of D1-14 of $\sim 1.4$, its erratic optical light curve, and the bell shape of its light curve in the NUV make this an unlikely tidal disruption event detection, and most likely a variable AGN.  The only remaining candidates that satisfy our criteria for a monotonic UV/optical flare and an inactive galaxy host are thus D1-9 and D3-13.

\section{Tidal Disruption Flare Candidates\label{sec:flare}}

\subsection{D1-9: Host Galaxy Properties}

Figure \ref{fig:spec} shows the spectrum of D1-9, which is well fitted by a Bruzual-Charlot early-type galaxy template with an old stellar population and no star formation at $z=0.326$.  The galaxy host is not detected in the FUV or NUV, so we can only put lower limits on its rest-frame $NUV-r$ color, which is a good diagnostic for separating galaxies into the red (early-type, old, $M> 3\times10^{10} \msun$) and blue (late-type, young, $M<3\times10^{10} \msun$) sequences \citep{kauff03}.  The lower limit on the UV-optical color, $NUV-r > 4$, places
the galaxy on the red side of the dividing line between the red and blue sequences \citep{wyd07}.  The optical colors of the host, which correspond to a rest-frame $u-r$=2.1 and $M_{r}=-20.9$, are close to the peak of the red sequence in the bimodal distribution of $u-r$ colors \citep{bald04}.  

The \textsl{XMM} Medium Deep Survey (XMDS) did not detect an X-ray source at the position of the host galaxy on 2002 July 25, placing an upper limit on its unabsorbed flux of $\aplt 1 \times 10^{-15}$ ergs s$^{-1}$ cm$^{-2}$ from 0.5 to 2 keV and $\aplt 7 \times 10^{-15}$ ergs s$^{-1}$ cm$^{-2}$ from 2 to 10 keV, for a power-law index ($f_{\nu} \propto \nu^{-(\Gamma-1)}$) typical of an AGN spectrum of $\Gamma=1.7$ and a Galactic column density towards the source of $N_{H}=2.61 \times 10^{20}$ cm$^{-2}$ \citep{chia05}.  For a luminosity distance of $d_{L}=1680$ Mpc, this corresponds to upper limits on the X-ray luminosity of $L_{X}$ (0.5-2 keV)$\aplt 3 \times 10^{41}$ ergs s$^{-1}$ and $L_{X}$ (2-10 keV)$\aplt 2 \times 10^{42}$ ergs s$^{-1}$.  

\subsection{D1-9: UV/Optical Light Curve}

\begin{figure}
\epsscale{.9}
\plotone{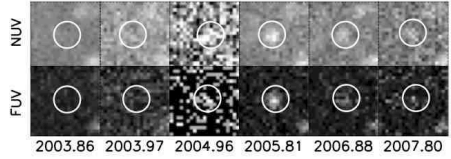}
\caption{\textsl{GALEX} images of the tidal disruption flare candidate D1-9 over 5 yr of DIS eclipses co-added into yearly epochs of NUV and FUV observations, labeled with the mean date of each coadd.  No source is detected in the NUV and FUV in the 2003.86 and 2003.97 co-added images.  Circles show the 6.0$\arcsec$ radius aperture used to measure the magnitudes.  \label{fig:fl}}
\end{figure}

\begin{figure}
\epsscale{.9}
\plotone{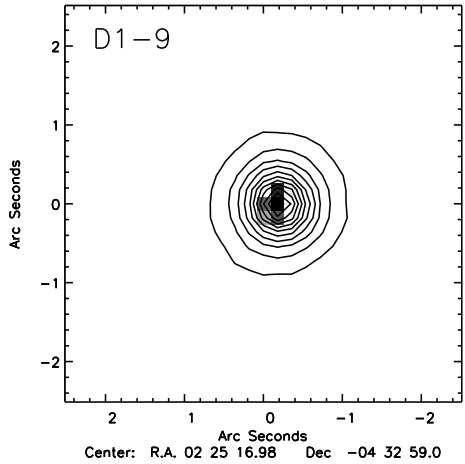}
\caption{Grey scale image of the variable optical point source shown on top of the $i$-band contours of the host galaxy of D1-9 from the CFHTLS Deep Imaging Survey. The variable optical source is coincident with the nucleus within the positional uncertainties of $\sim 0\farcs25$. \label{fig:cont}}
\end{figure}

\begin{figure}
\epsscale{.9}
\plotone{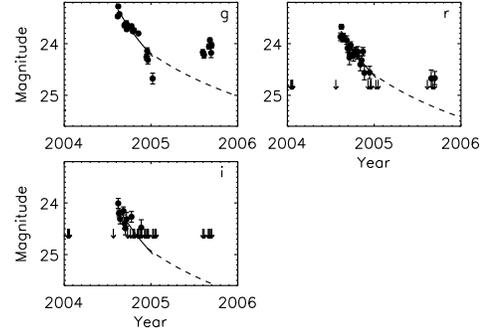}
\caption{Optical light curve of the tidal disruption flare candidate D1-9 measured from difference imaging by the CFHTLS.  Error bars show 1 $\sigma$ errors, and arrows show upper limits based on the 5 $\sigma$ detection of a point source.  The optical photometry catches the steep rise of the flare to its peak on 2004 August 14.  The solid line shows the  ($t-t_{D})^{-5/3}$ fit to the first 5 months of the $g$-, $r$-, and $i$-band light curves, for $g-r=-0.41$ and $r-i=-0.33$, and the dashed line shows the best-fitting power-law decay for the NUV data with $t_{D}$ fixed to 2004.0.\label{fig:opt}}
\end{figure}

Figure \ref{fig:fl} shows the FUV and NUV yearly coadded images of the tidal disruption flare candidate D1-9.  We include a co-added image from 2007 with $t_{\rm exp}=8.2$ ks that was obtained after the selection of candidates, and is not listed in Table 1.  The source is undetected in the FUV and NUV in co-added images in 2003 composed of visits from 2003 October 27 to November 23 and visits from December 20 to 22.  The source appears in 2004 December 16, indicating an amplitude of variability of $\Delta m \apgt$ 2 mag.  The source then decays monotonically by $\sim 1.0$ mag over 3 yr to below the detection threshold in 2007.  The UV flare from D1-9 was also detected as an optically variable source in the nucleus of the galaxy by the CFHTLS.  Figure \ref{fig:cont} shows a gray-scale image of the variable optical source on top of the contours of the host galaxy in the $i$ band.  Figure \ref{fig:opt} shows the optical light curve of the flare, with the flux from the host galaxy subtracted off,  in the CFHTLS $g, r, $ and $i$ bands.  The variable optical source is undetected in the $z$ band, with $z > 23.2$.

\begin{figure}
\epsscale{.9}
\plotone{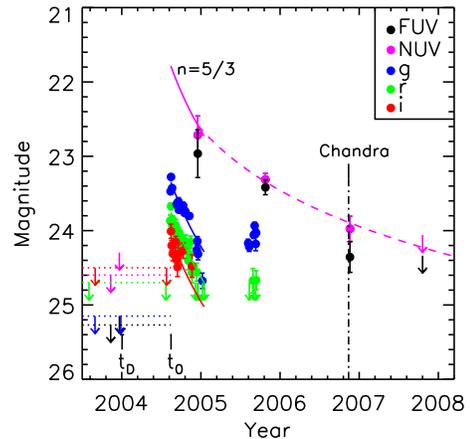}
\caption{UV/optical light curve of the tidal disruption flare candidate D1-9 measured by \textsl{GALEX} and CFHTLS.  Error bars show 1 $\sigma$ errors, and solid arrows show upper-limits.  The least-squares fit to the first 5 months of the optical light curve of the flare with a ($t-t_{D})^{-n}$ decay with $n=5/3$
is shown with a thick solid line.  The dashed line shows the best-fitting power-law index to the NUV data with $t_{D}$ fixed to 2004.0, with $n=1.1 \pm 0.3$ and with NUV$-r=-2.0\pm$0.2.  The dot-dash line
indicates the time of a 30 ks ToO $0.2-10$ keV \textsl{Chandra} observation that detected a soft X-ray source. \label{fig:lc}}
\end{figure}

The $r$-band photometry catches the steep rise of the flare from a non-detection from 2003 August 1 to 2004 July 24 to its peak on 2004 August 14.  Two $r$-band data points were removed from
2005 August 4 and 2005 September 4 because of systematic errors introduced by a poor convolution kernel that under subtracted the galaxy near the nucleus, and results in a variable source flux that is systematically too bright.  Figure \ref{fig:lc} shows the simultaneous UV/optical light curve of the flare.  Due to the small number of counts, the FUV and NUV 1 $\sigma$ error bars and 95\% confidence
upper limits are measured using Bayesian statistics.  

We fit the $g$-, $r$-, and $i$-band light curves simultaneously for $g-r=-0.41$ and $r-i=-0.33$ with a $(t-t_{D})^{-n}$ power-law decay for data points from the peak of the flare to 2005.04.  For $n=5/3$, the power-law expected for a tidal disruption event, this results in a least-squares fit to the time of disruption, $t_{D}=2004.00 \pm 0.02$ (reduced $\chi^{2}$=2.8).  If we allow the power-law index $n$ to vary, we get $t_{D}=2003.2 \pm 0.2$ and $n=3.4 \pm 0.4$ (reduced $\chi^{2}$=2.8).  There is a large difference in $t_{D}$ for the different power-laws, but with the same goodness-of-fit. 

The excellent temporal sampling of the CFHTLS observations measures the time of the peak of the flare, $t_{0}=2004.622$, and we can determine the rest-frame time delay of the most tightly bound material to return to pericenter to high accuracy.  If we assume the more physically motivated $n=5/3$ fit to the light curve, then $(t_{0}-t_{D})/(1+z) = 0.47 \pm 0.02$ yr, which implies $\mbh = (1.8 \pm 0.2) k^{3} \times 10^{7} \msun$ for a solar-type star disrupted at $R_{T}$, where $k$ depends on the spin-up of the star on disruption.   If we use the time of disruption from the $n=3.4$ fit, this yields
a larger time delay, $(t_{0}-t_{D})/(1+z) = 1.1 \pm 0.2$ yr, and a much higher black hole mass, $\mbh = (1.0 \pm 0.4) k^{3} \times 10^{8} \msun$, which will quickly exceed $M_{crit}$ for stars with some spin on disruption (see \S\ref{sec:theory}).

Although the first 5 months of the flare light curve, from 2004.622 to 2005.04, are well described by a $(t-t_{D})^{-5/3}$ power-law decay, optical photometry in the $g, r$ and $i$ bands a year after the peak of the flare and the UV light curve extending 3 yr after the flare indicate that the emission does not continue to decay with the same power-law, but instead with a shallower decay.  We fit the NUV light curve with a power-law decay with $t_{D}$ fixed to 2004.0 and find a best-fitting power-law index of $n=1.1\pm0.3$ and NUV$-r=-2.0 \pm 0.2$. 
This power-law in the UV light curve is not consistent with the plateau in the optical data in late 2005 and may indicate a change in the UV/optical colors after 2005.04 closer to $NUV-r \sim -0.6$.
The dip in the $g$-band light curve on 2005.04 is not seen in the other optical bands and may be a measurement that is below the detection threshold of the observations.  It may also be a systematic error, since the errors in the CFHTLS difference imaging fluxes are measured from the dispersion of flux measurements on a given night and do not account for systematic errors that can occur from constructing the difference image.  If the dip is real, then it is a significant deviation from a simple power-law decay.

\subsection{D1-9: Spectral Energy Distribution}

\begin{figure}
\epsscale{.9}
\plotone{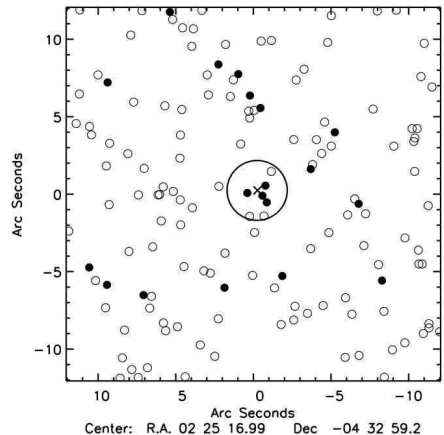}
\caption{\textsl{Chandra} ACIS-S image of the tidal disruption flare candidate D1-9 on 2006 November 12.  Photons with energies $>$ 1 keV are plotted with open circles, and photons with energies $<$ 1 keV are plotted with filled circles.  There is a statistically significant detection of four soft photons with energies between 0.2 and 0.4 keV that are detected at the position of the UV flare (labeled with a cross). The circle shows the aperture ($r_{ap}$=4 physical pixels) used to measure the count rate.  \label{fig:xray}}
\end{figure}

We triggered a 30 ks \textsl{Chandra} ACIS-S (0.2--10 keV) Target of Opportunity (ToO) X-ray observation of the flare on 2006 November 12 and detected four soft photons with energies between 0.2 and 0.4 keV (shown in Figure \ref{fig:xray}).  The background counts with energies $<$1 keV have a mean of 0.034 counts arcsec$^{-2}$.  For a Poisson distribution with a mean count rate within $r_{ap}=1.968\arcsec$ of 0.4 counts, the probability of having 4 counts from the background is $< 0.07$\%, and so we consider this source a statistically significant detection.  Our observation also puts an upper limit on the unabsorbed hard X-ray luminosity, L(2--10 keV) $< 6 \times 10^{41}$ ergs s$^{-1}$, for an AGN power-law $\Gamma = 1.7$ and $N_{H}=2.61 \times 10^{20}$ cm$^{-2}$.

\begin{figure}
\epsscale{.9}
\plotone{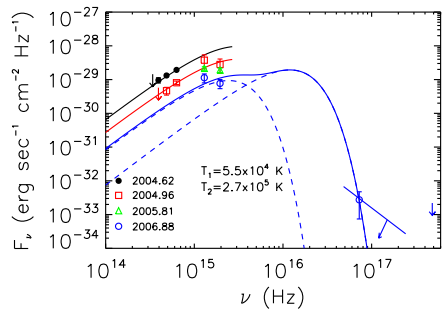}
\caption{SED measured by CFHTLS, \textsl{GALEX}, and \textsl{Chandra} of the tidal disruption flare candidate D1-9 over time.  Fluxes have been corrected for Galactic extinction of
$N_{H}=2.61 \times 10^{20}$ cm$^{2}$, and error bars show 2 $\sigma$ errors.  Solid lines show the best fitting two-temperature blackbody spectrum, and the dashed line shows the individual blackbody components.  The $\Gamma > 3$ power-law spectrum of the soft X-ray detection and the upper limit to the hard X-ray flux density at 2 keV are also shown.  \label{fig:sed}}
\end{figure}

We plot the SED of the flare over time from the optical through the X-rays in Figure \ref{fig:sed}.  The UV/optical fluxes have been corrected for Galactic extinction using the extinction curve of Cardelli \etal (1989).  The simultaneous UV/optical fluxes on 2004.96 are well fitted by a blackbody with a rest frame $T_{bb}=(5.5 \pm 1.0) \times 10^{4}$ K and an $r$-band flux of $(4.7 \pm 0.6) \times 10^{-30}$ ergs s$^{-1}$ cm$^{-2}$ Hz$^{-1}$.  Using $L_{\nu_{e}=(1+z)\nu_{o}}=  f_{\nu_{o}} (4\pi d_{L}^2)/(1+z) = (1.2 \pm 0.2) \times 10^{27}$ ergs s$^{-1}$ Hz$^{-1}$ in the $r$ band and $L_{\nu}=(2\pi h/c^{2})\nu^{3}/(e^{h\nu/kT}-1)4\pi R_{bb}^2$, this corresponds to $R_{bb}=(7.6 \pm 0.6) \times 10^{13}$ cm and $L_{bol}=\sigma T^{4} 4\pi R_{bb}^2 = (3.8 \pm 0.7) \times 10^{43}$ ergs s$^{-1}$.  If we assume that the temperature of the emission does not change, and we scale this blackbody to the optical fluxes at the flare's peak on 2004.62, this corresponds to a bolometric luminosity of $(9.1 \pm 0.2) \times 10^{43}$ ergs s$^{-1}$.  
When we fit the optical fluxes at the peak with a power-law, $F_{\nu} \propto \nu^{\alpha}$, we get $\alpha = +1.5 \pm 0.01$, which is unlike the $\alpha <-.53$ typical of AGNs and is better ascribed to the Rayleigh-Jeans tail of a blackbody spectrum.

During the soft X-ray detection on 2006.88 we require a second higher temperature blackbody to fit the X-ray flux density at 0.3 keV, with $T_{bb}=(2.7 \pm 0.2) \times 10^{5}$ K.  For this blackbody spectrum, the soft X-ray count rate on 2006.88 corresponds to an unabsorbed (0.2-1.0 keV) luminosity of $(3.5 \pm 0.2) \times 10^{41}$ ergs s$^{-1}$.  This blackbody has
$R_{bb} = (6 \pm 1) \times 10^{12}$ cm and  $L_{bol} = (1.4 \pm 0.6) \times 10^{44}$ ergs s$^{-1}$.
Due to the small number of counts, we cannot fit the spectrum of the X-ray emission, and so we cannot be sure that it is truly a soft blackbody.  However, we can place limits on a power-law fit, $F_{E} \propto E^{-\Gamma}$, to the spectrum based on the lack of a source detection above 1 keV, to have $\Gamma > 3$ and $L(0.2-1.0$ keV) $\aplt 5 \times 10^{41}$ ergs s$^{-1}$.
Unfortunately, we do not have X-ray observations during the peak of the flare to determine if the high-temperature component is present and with the same ratio to the lower temperature blackbody as at later times. 

\subsection{D3-13: Host Galaxy Properties}

D3-13 was reported in \citet{gez06} for having an AEGIS \citep{dav06} Keck DEIMOS spectrum that classified the host galaxy as an inactive early-type galaxy at $z=0.3698$.  We also obtained a Keck LRIS spectrum of D3-13 (shown in Figure \ref{fig:lris2}) to cover the wavelength region near H$\alpha$, and no H$\alpha$ emission is detected, consistent with its previous classification as an inactive galaxy.  
The galaxy host is not detected in the FUV or NUV, so we can only put lower limits on its rest-frame NUV$-r \apgt 4$, which places the galaxy on the red side of the dividing line between the red and blue sequences \citep{wyd07}.  The optical colors of the host galaxy, which correspond to a rest-frame $u-r$=2.0 and $M_{r}=-21.7$, place it on the blue end of the $u-r$ color distribution corresponding to the red sequence \citep{bald04}.  The location of the galaxy on the edge of the red sequence is in agreement with the morphology of the galaxy, since the bulge-disk composition of its AEGIS \textsl{HST} ACS $I$-band image
does detect a disk component, with a bulge fraction of 0.72 \citep{gez06}.  No hard X-ray source was detected in deep (190 ks) archival AEGIS \textsl{Chandra} ACIS-I observations of the source taken over 2005 April -- September, resulting in an upper limit to the unabsorbed ($2-10$ keV) X-ray luminosity of $< 1.5 \times 10^{41}$ ergs s$^{-1}$, for an AGN power-law $\Gamma = 1.4$ and $N_{H}=1.3 \times 10^{20}$ cm$^{-2}$, and $d_{L}=1970$ Mpc.   

\subsection{D3-13: UV/Optical Light Curve}

\begin{figure}
\epsscale{.9}
\plotone{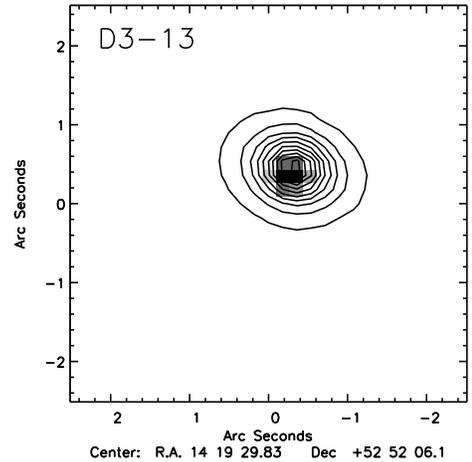}
\caption{Gray-scale image of the variable optical point source shown on top of the $i$-band contours of the host galaxy of D3-13 from the CFHTLS Deep Imaging Survey. The variable optical source is coincident with the nucleus within the positional uncertainties of $\sim 0\farcs25$. \label{fig:cont2}}
\end{figure}

\begin{figure}
\epsscale{.9}
\plotone{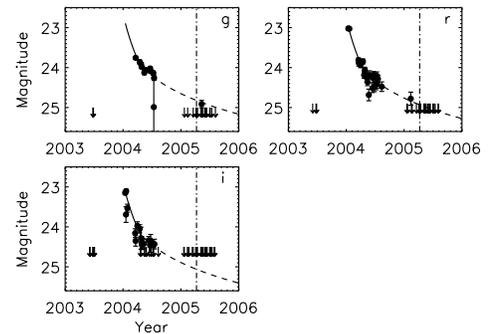}
\caption{Optical light curve of the tidal disruption flare candidate D3-13 measured from difference imaging by the CFHTLS.  Error bars show 1 $\sigma$ errors, and arrows show upper limits based on the 5$\sigma$ detection of a point source.  The $r$- and $i$-band photometries catch the light curve close to the peak of the flare.  The solid line shows the  ($t-t_{D})^{-5/3}$ fit to the $g$-, $r$-, and $i$-band light curves for $g-r=-0.11$ and $r-i=-0.13$, and the dashed line shows the best-fitted power-law index for the NUV data with $t_{D}$ fixed to 2003.7. The dot-dash line indicates the time of the soft X-ray detection by \textsl{Chandra}.\label{fig:opt_groth}}
\end{figure}

The UV light curve of D3-13 was presented in \citet{gez06}, with no optical variability reported in CFHTLS data spanning from 2005 January to June with $m_{lim} \sim 25$.  Here we extend the UV light curve with one more epoch of \textsl{GALEX} DIS data from 2007.  We also use a deeper co-added image in 2004 with $t_{exp}=8.3$ ks, which includes visits that have positional offsets from the center and thus were not included in the main coadd.  We also present new data from CFHTLS that detect an optically variable source in the nucleus of the host galaxy in 2004 (Figure \ref{fig:cont2}).  Figure \ref{fig:opt_groth} shows the optical light curve in the $g$, $r$, and $i$ bands with the flux from the host galaxy subtracted off.  The variable source is undetected in the $z$ band, with $z > 23.2$.  

The $r$- and $i$-band light curves catch the flare close to its peak on 2004.04 and its subsequent steep decay.  The light curves are best described with a broken power-law.  A solid line shows the best-fitted $n=5/3$ power-law decay to the first few months of the $g$-, $r$-, and $i$-band data from 2004.04 to 2004.35, simultaneously fitted for $g-r=-0.11$, $r-i=-0.13$, yielding a tight constraint on $t_{D}=2003.71 \pm 0.01$ (reduced $\chi^{2} = 3.2$).  When we allow $n$ to vary, the resulting fit yields $t_{D}=2003.686 \pm 0.003$ and $n=1.77 \pm 0.01$ (reduced $\chi^{2} = 2.9$).  We include the sensitivity of the time of disruption to the power-law index in the error bars for $t_{D}$ and use $t_{D}=2003.70 \pm 0.02$.  

\begin{figure}
\epsscale{.9}
\plotone{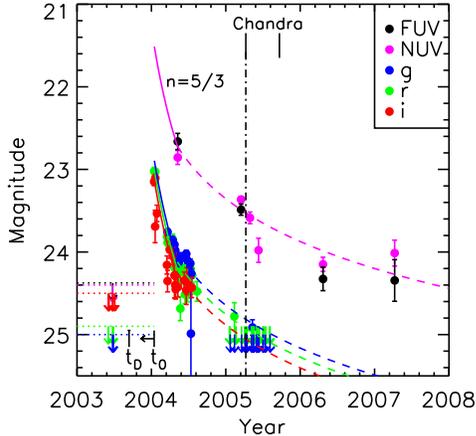}
\caption{UV/optical light curve of the tidal disruption flare candidate D3-13 measured by \textsl{GALEX} and CFHTLS.  Error bars show 1 $\sigma$ errors, and solid arrows show upper-limits.  The least-squares fit to the first 4 months of the optical light curve of the flare with a ($t-t_{D})^{-n}$ decay with $n=5/3$
is shown with a thick solid line.  The best-fit power-law index for the NUV data after 2004.04 with $t_{D}$ fixed to 2003.7 and NUV$-r=-1.4$ is shown with a dashed line with $n=0.82 \pm 0.03$ .  Tick marks indicate the time of archival AEGIS 100 ks $0.2 - 10$ keV \textsl{Chandra} observations.  A dot-dash line indicates the time of a soft X-ray detection by \textsl{Chandra}.  \label{fig:lc_groth}}
\end{figure}

Figure \ref{fig:lc_groth} shows the simultaneous FUV, NUV, and optical light curve.  Due to the small number of counts, the FUV and NUV 1 $\sigma$ error bars and 95\% confidence
upper limits are measured using Bayesian statistics.  The NUV photometry was measured with $r_{ap}=3.75\arcsec$ and $f_{ap}=0.70$ to avoid contamination by a pair of foreground-interacting galaxies detected in the NUV just south of the source.  For the later observations in 2005 the FUV detector was temporarily not operational.  

The UV data trace the decay of the flare at later times, which has a shallower power-law.  
In \citet{gez06} we fitted the NUV light curve with an $n=5/3$ power-law decay, and found a best fit $t_{D}=2003.3 \pm 0.2$.  When the power-law index was allowed to vary, we found $t_{D}=2002 \pm 2$ and $n=3 \pm 2$.  
Here we fit the optical and NUV data points after 2004.35 with NUV$-r=-1.4$, the optical colors from the $n=5/3$ fit, and with $t_{D}$ fixed to 2003.7, and we get a best-fit power-law index of $n=0.82 \pm 0.03$.    

Although the optical light curve of D3-13 puts strong constraints on $t_{D}$, we do not see the rise of the flare to its peak, and so we use the earliest detection of the flare in the $r$ and $i$ band on 2004 January 13 to place an upper limit on ($t_{0}-t_{D})/(1+z) < 0.25$ yr, which corresponds to $\mbh < $5.2 $\times 10^{6} k^{3} \msun$ for a solar-type star that is disrupted at $R_{T}$.  

\subsection{D3-13: Spectral Energy Distribution}

\begin{figure}
\epsscale{.9}
\plotone{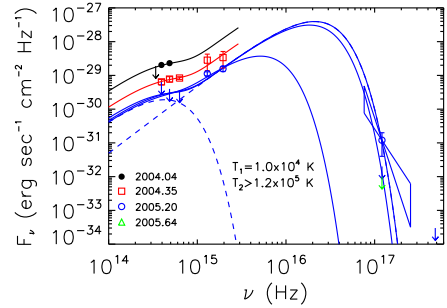}
\caption{SED measured by CFHTLS, \textsl{GALEX}, and \textsl{Chandra} of the tidal disruption flare candidate D3-13 over time.  Fluxes have been corrected for Galactic extinction of $N_{H} = 1.30 \times 10^{20}$ cm$^{-2}$, and error bars show 2 $\sigma$ errors.   Solid lines show the best-fitting two-temperature blackbody spectrum, and the dashed line shows the individual blackbody components.  The steep slope of the soft X-ray detection ($\Gamma = 7 \pm 2$) and the upper limit to the hard X-ray flux density at 2 keV are also shown.  \label{fig:sed_groth}}
\end{figure}

We plot the broadband SED of D3-13 over time from the optical through the X-rays in Figure \ref{fig:sed_groth}.  Two archival 45 ks AEGIS \textsl{Chandra} ACIS-I observations were taken on 2005 April 6 and 7, and an extremely soft source with 10 photons with energies between 0.3 and 0.8 keV appeared in the first $\sim 30$ ks of the second day of observations, with a steep spectral slope of $\Gamma=7 \pm 2$ and a corresponding unabsorbed luminosity of $L(.3-1$ keV) $\sim 1 \times 10^{43}$ ergs s$^{-1}$.  During the 45 ks observation on 2005 April 6, and during a later 100 ks observation on 2005 September 20-23, no source was detected from 0.3 to 2.0 keV.  For a source with the same spectrum, this places upper limits on $L(0.3-1$ keV) of $< 2 \times 10^{42}$ and $< 1 \times 10^{42}$ ergs s$^{-1}$, indicating variability of a factor of 10 on the timescale of 1 day. The variable soft X-ray source is unlikely a contaminant in the image, since the duration of the flare is much longer than a cosmic-ray afterglow, which typically lasts on the order of hundreds of seconds (E.~Laird 2007, private communication).  

The simultaneous UV/optical SED of the flare on 2004.35 requires a two temperature blackbody fit.  The low-temperature ($T_{1}$) blackbody component has a rest-frame $T_{bb}=(1.0 \pm 0.1) \times 10^{4}$ K, with $R_{bb}= (4.7 \pm 0.3) \times 10^{14}$ cm and $L_{bol}= (1.6 \pm 0.7) \times 10^{42}$ ergs s$^{-1}$.  The higher temperature ($T_{2}$) blackbody spectrum is well constrained by the soft X-ray flux detection on 2005 April 7, with $T_{bb}= (4.9 \pm 0.2)\times 10^{5}$ K, and the steep slope of the soft X-ray detection could be ascribed to emission from the Wien's tail of a blackbody.  
This blackbody has $R_{bb}= (1.3 \pm 0.2) \times 10^{13}$ cm and $L_{bol}= (7 \pm 1) \times 10^{45}$ ergs s$^{-1}$.  The lowest temperature for $T_{2}$ that is consistent with the UV/optical fluxes on 2004.35 is $T_{bb}=1.2 \times 10^{5}$ K and corresponds to $R_{bb}= 4.7 \times 10^{13}$ cm and $L_{bol}> 3.3 \times 10^{44}$ ergs s$^{-1}$.  

The bolometric luminosity of the flare is dominated by the higher temperature blackbody component.  If we scale the blackbody model to the optical fluxes near the peak of the flare on 2004.04, this indicates a peak bolometric luminosity (for $T_{2}> 1.2 \times 10^{5}$ K) of $> 2.5 \times 10^{45}$ ergs s$^{-1}$.
If we fit the optical fluxes with a power-law, we get $\alpha = +0.5 \pm 0.2$, which is unlike the $\alpha <-.53$ typical of AGNs.

\section{Interpretation}

The UV--optical colors of the candidate flares (NUV$-r < -1$) are incompatible with SNd Ia and GRB afterglows, which are intrinsically faint in the UV, and their locations at the centers of the galaxies are not characteristic of SNe II which are associated with massive star formation in the disks of galaxies.  The peak bolometric luminosity of the nuclear flares ($> 10^{44}-10^{45}$ ergs s$^{-1}$) 
makes them most likely associated with an outburst
from a central SMBH with $\mbh > 10^{6} \msun$.  \textsl{Chandra} observations of the host galaxies put limits on the presence of hard X-ray emission from a power-law AGN in the 
nuclei of the galaxies down to $L_{X} \aplt 10^{41}$ ergs s$^{-1}$, and ground-based optical spectra put
limits on optical emission lines powered by an AGN to L([O~III]$\lambda5007) < 10^{40}$ ergs s$^{-1}$. 
These upper limits do allow for the presence of a low-luminosity AGN (LLAGN) in these galaxies at the faint end of the luminosity function measured for nearby LLAGNs \citep{heck05}. 
In addition, the amplitude of variability in the UV in these galaxies ($\Delta m > 2$) is comparable to that of some of the AGNs and quasars in our sample.  However, the detailed optical light curves of these candidates, with their steep rise to the peak and subsequent power-law decay, are distinctive from AGNs, as are the positive optical power-law slopes of their SEDs ($\alpha \ge +0.5$).  The combination of the power-law light curves, blackbody SEDs, soft X-ray properties, lack of detectable Seyfert activity, and large luminosities make a tidal disruption event the most likely explanation for the source of the UV/optical flares.

\begin{figure}
\epsscale{.9}
\plotone{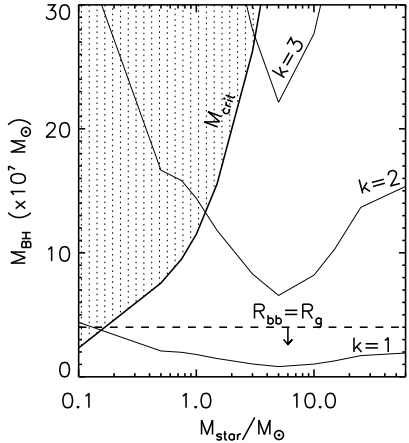}
\caption{Solutions for the central black hole mass of D1-9 as a function of mass for main sequence stars with a spin-up parameter $k$.  The thick solid line shows the maximum possible black hole mass for $R_{T}>R_{Sch}$, and the dashed line shows the maximum black hole mass for $R_{bb}>R_{g}$. \label{fig:plot_mbh}}
\end{figure}

The change in the exponent of the power-law decay in the light curves of the candidates occurs ($t-t_{D}$)/(1+$z$) = 0.78 and 0.47 yr after the disruption for D1-9 and D3-13, respectively.  These timescales are consistent with the timescale for circularization of the debris, which is expected to be on the order of the timescale of the minimum period of the orbital debris.  This minimum period is measured for D1-9 and D3-13 to be ($t_{0}-t_{D}$)/(1+$z$) = 0.47 and 0.25 yr, respectively, which would correspond to circularization in $\aplt 2$ orbital periods.  Numerical simulations from \citet{ayal00} also show that the accretion rate can also deviate from a $(t-t_{D})^{-5/3}$ power-law during the fallback phase due to the expulsion of debris from compression during the second passage through pericenter.  The
broken power-law model does not describe the plateau in the optical light curve (and dip in the $g$ band) of D1-9, or the nearly constant UV flux of D3-13 at late times.  The emission mechanisms during a tidal disruption event (fallback, stream-stream collisions, compression, accretion) are complicated and not well understood.
We are seeing the messy details of the light curves of these events for the first time, however, and it is impressive that a simple broken power-law model reproduces the observations as well as it does.

\begin{figure}
\epsscale{.9}
\plotone{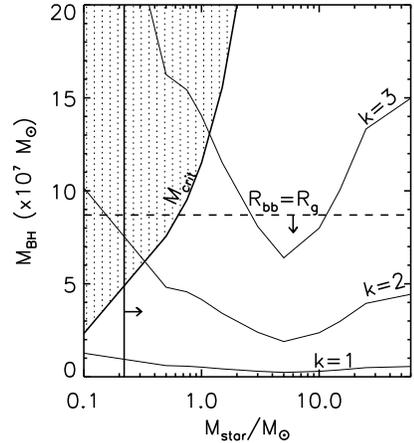}
\caption{Solutions for the central black hole mass of D3-13 as a function of mass for main sequence stars with a spin-up parameter $k$.  The thick solid line shows the maximum possible black hole mass for $R_{T}>R_{Sch}$, and the dashed line shows the maximum black hole mass for $R_{bb} > R_{g}$. The horizontal arrow shows the minimum mass of the disrupted star derived from the minimum mass accreted.  \label{fig:plot_mbh_groth}}
\end{figure}

\subsection{Central Black Hole Mass}

If we assume that the debris has shocked and circularized into an accretion torus or disk after the change in
the power-law of the light curve, then the soft X-ray emission detected during this time may originate from the inner regions of the disk or torus.  This interpretation would also account for the rapid variability observed for the soft X-ray source in D3-13.  The radius of emission during the soft X-ray detections can place constraints on the central black hole mass, since this emission should have a pericenter radius that is greater than the minimum stable particle orbit for the black hole ($R_{ms}$), which ranges from $R_{ms} = 6R_{g}$ for a black hole with no spin down to $R_{ms}$ = $R_{g}$ for a maximally spinning black hole, where R$_{g} = G\mbh/c^{2}$.  If we require that $R_{bb} > R_{ms}$, then the blackbody radii of the soft X-ray components for D1-9 and D3-13 of $6 \times 10^{12}$ and $1.3 \times 10^{13}$ cm, respectively, require that $\mbh < 4 \times 10^{7} \msun$ and $8.7 \times 10^{7} \msun$, which are also consistent with $\mbh < M_{crit}$ for a solar-type star.

There is an uncertainty in the peak bolometric luminosity of the flares in D1-9 and D3-13 due to the lack of X-ray observations close to the peak of the flare.  However, we can estimate the minimum mass accreted during the fallback phase of the debris by taking the lower limit to the peak luminosity,
and integrating under the portion of the light curve well described by a $t^{-5/3}$ power-law to get the total energy output.   The luminosity of the lower temperature component of D1-9 gives a lower-limit to its peak bolometric luminosity of $9.1 \times 10^{43}$ ergs s$^{-1}$.  We integrate under the portion of the light curve well described by a $t^{-5/3}$ power-law to get
$\int _{t_{0}}^{2005.04} L_{0} [(t-t_{D})/(t_{0}-t_{D})]^{-5/3} dt = 8.0 \times 10^{50}$ ergs s$^{-1}$ which corresponds to $E_{tot}/(\epsilon c^2) = 0.004 \msun$ accreted, for an efficiency of converting mass into energy of $\epsilon = 0.1$.  This places a lower limit to the mass of the disrupted star to be $0.008 \msun$ if at least half of the stellar debris is unbound from the system at the time of disruption.  If we do the same exercise for D3-13, using the lower-limit to the peak luminosity from the lower bound to $T_{2}$ of $2.5 \times 10^{45}$ ergs s$^{-1}$, and integrate from $t_{0}$ to 2004.35, this results in a minimum mass accreted of 0.11 $\msun$, which requires a star of at least $0.22 \msun$. 

In Figures \ref{fig:plot_mbh} and \ref{fig:plot_mbh_groth} we plot the solutions for $\mbh$ from the rest-frame time delay between the time of the disruption and the peak of the flare, $(t_{0}-t_{D})/(1+z)$, as a function of main-sequence star mass and $k$, in comparison to $M_{crit}$, which is the upper limit to the black hole mass for which a star will be disrupted, i.e. $R_{T} > R_{Sch}$.    For $1<k<3$, i.e. the star is spun-up to some fraction of break-up, we get a range of allowable solutions for $\mbh$ for D1-9 from $8.2 \times 10^{6} \msun$ to the upper limit from $R_{bb}>R_{g}$ of $4 \times 10^{7} \msun$.  The maximum allowable value for $k$ in this range of black hole masses is $k= 1.7$.  For D3-13 the full range of $k$-values is allowed,  which yields a large range of black hole masses from a minimum of $2.4 \times 10^{6} \msun$ to the upper limit from $R_{bb}>R_{g}$ of $8.7 \times 10^{7} \msun$. 

We can compare these values for $\mbh$ with the mass expected from the properties of the host galaxies.  For D1-9 the rest-frame absolute $B$-band Vega magnitude is M$_{B}= g-5\log(d_{L})+5-K = -19.5$, where $d_{L}$ is in Mpc and $K$ = 0.68.  This can be related to black hole mass using the correlation between $\mbh$ and $L_{B}$(bulge) from \citet{mag98} which has a large scatter ($\sim$0.5 dex) that yields $\mbh \sim 1^{+2}_{-.7} \times 10^{8} \msun$, which is compatible with solutions in Figure \ref{fig:plot_mbh}.
Because of the high spectral resolution of the AEGIS Keck DEIMOS spectrum of D3-13, it is possible to measure the velocity dispersion ($\sigma$) of the bulge directly and use the observed tight correlation between $\mbh$ and $\sigma$ \citep{geb00}.  The $\mbh$ estimated from the velocity dispersion, measured to be $\sigma_{\star}=120 \pm 10$ km s$^{-1}$, is 2$^{+2}_{-1} \times 10^{7} \msun$ \citep{gez06} and is compatible for solutions in Figure \ref{fig:plot_mbh_groth} with $k \aplt 2$.

\section{Tidal Disruption Rate \label{sec:rate}}
In this section we describe how we measure the efficiency of our search over 2.882 deg$^{2}$ of DIS and simulate the tidal disruption flare detection rate expected from theoretical models .  We use the tidal disruption rates from \citet{wm04} calculated for 51 elliptical galaxies with measured surface brightness profiles.  We fit the galaxies with $\mbh \aplt 10^{8} \msun$ with a power-law to get a black-hole-mass-dependent tidal disruption rate $\dot{N}(\mbh) \sim 1.6 \times 10^{-4} yr^{-1} (\mbh/10^{6} \msun)^{-.30}$.  This gives lower rates than the analytically determined rate for a singular isothermal sphere with a stellar density profile of $\rho \propto r^{-2}$ =  $\dot{N}(\mbh) \sim 6.5 \times 10^{-4} yr^{-1} (\mbh/10^{6} \msun)^{-.25}$.   We assume that the flares radiate at the Eddington luminosity,

\begin{figure}
\epsscale{.9}
\plotone{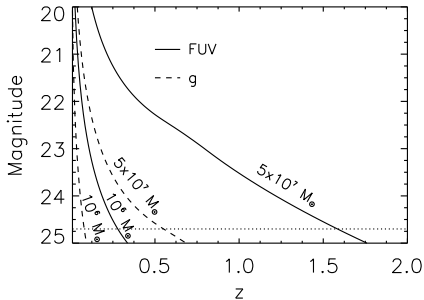}
\caption{Predicted peak FUV and $g$ magnitudes and of flares from $10^{6} - 5 \times 10^{7} \msun$ black holes in comparison to the detection limit for a 20 ks \textsl{GALEX} DIS exposure and a 720 s Pan-STARRS MDS exposure.  \label{fig:app}}
\end{figure}

\begin{equation}
$$\[L_{\rm Edd} = 1.3 \times 10^{44} (\mbh/10^{6} \msun) {\rm ergs} {\rm~s}^{-1},\]$$
\end{equation}

\noindent and with an effective temperature characterized by Eddington luminosity radiation at the tidal disruption radius,

\begin{equation}
$$\[T_{\rm eff} = 2.5 \times 10^{5} (\mbh/10^{6} \msun)^{1/12} {\rm~K}.\]$$
\end{equation}

\noindent  Similarly to the calculation of \citet{mt99}, we derive the black hole mass function from the
\citet{ferg91}
E + S0 luminosity function,

\begin{equation}
$$\[N(L){\rm d}\log{L} = \frac{N_{0}}{(\sqrt{2\pi} \Delta)}\exp\left[-\frac{1}{2}\left(\frac{\log{L}-\log{L_{0}}}{\Delta}\right)^{2}\right] {\rm d}\log{L},\]$$
\end{equation}

\noindent where $\Delta=0.6$, $N_{0}=3.0 \times 10^{-3}$ Mpc$^{-3}$, and $L_{0}=2.6 \times 10^{9} L_{\sun}$, with a factor of 2 to account for bulges in early-type spirals,
scaled to black hole mass by combining the $\mbh$ - M$_{bulge}$ relation from \citet{merr01}
and the mean galaxy mass-to-light ratio from \citet{mag98} which yields,
$\mbh = 1.6 \times 10^{-4} \left(\frac{L_{bulge}}{L_{\odot}}\right)^{1.18} \msun$.
We only consider the range of black hole masses for which a tidal disruption flare can
radiate close to L$_{Edd}$, the 
10$^{6} - 5 \times 10^{7} \msun$ mass range \citep{ulmer99}.  Figure \ref{fig:app} shows the predicted FUV magnitudes of flares from $10^{6} - 5 \times 10^{7} \msun$ black holes in comparison to the detection limit for a 20 ks DIS exposure.  We include an estimate of the attenuation in the FUV band by neutral hydrogen absorption systems in the line of sight from \citet{mad95}, which produces a non-negligible effect for sources at $z > 0.6$.  The $K$-correction,
$K(z) \approx$ (-7.5)log(1+z), is large, and thus Eddington luminosity flares from a 5 $\times$ 10$^{7} \msun$ black hole can be detected in the FUV band out to $z\sim1.5$.  

The detectability of a flare by \textsl{GALEX} also depends on the FUV magnitude of its host galaxy.   We estimate the FUV magnitude of the host galaxy of a given black hole mass by converting the $B$-band luminosity of the galaxy to a $g$-band apparent magnitude using average $g-B$ corrections from \citet{fuk95} measured for elliptical/S0 galaxies of $-0.375$ and for Sab/Sbc spiral galaxies of $-0.340$.  We then convert the $g$ magnitude to an FUV magnitude with $g-$FUV$ > 3$ for elliptical/S0 galaxies and $g-$FUV$ > 2$ for Sab/Sbc spiral galaxies.  The number of events detected within a redshift $z$ over a baseline of observations of $\Delta t$ years is thus

\begin{eqnarray*}
N(< z) = A\Delta t~\times \hspace{2.5in}  & \\
 \int_{10^6 \msun}^{5 \times 10^7 \msun} \dot{N}(\mbh) N_{BH}(\mbh) V(z) S(\mbh,z) f(\mbh) {\rm d}\mbh, & 
\end{eqnarray*}

\noindent where $A$ is the area in degrees, $S(\mbh,z)$ is the selection function, and $f$ is the fraction of flares that radiate at the Eddington luminosity for a solar-type star from \citet{ulmer99}.  Because the peak accretion rate of the debris is proportional to $\mbh^{-1/2}$, as the central black hole mass increases, the fraction of flares that have a mass accretion rate close to the Eddington limit and will have $L_{peak} \approx L_{\rm Edd}$ decreases.  We use $f=1$ for $\mbh < 4 \times 10^{6} \msun$, and for $\mbh \ge 4 \times 10^{6} \msun$ we use the log-linear function,

\begin{equation}
$$\[f(\mbh) = -0.909 \log(\mbh/\msun)+7.00.\]$$
\end{equation}

\noindent We estimate the selection function with a Monte Carlo simulation.  We simulate the number of flares predicted for each black hole bin for the redshift volume $V(z)$, selected from a Poisson distribution with a mean $\langle N(\mbh, z )\rangle =A\Delta t \dot{N}(\mbh) N_{BH}(\mbh) V(z) f(\mbh)$, and randomly select $z$ and $t_{0}$ for each flare.  Values of $t_{0}$ are uniformly distributed between 2003 and 2003+$\Delta{t}$, and the distribution of $z$-values is proportional to the co-moving volume at each redshift,  $V(z)=\frac{4}{3}\pi \left[\frac{d_{L}}{(1+z)}\right]^3\frac{\theta}{41,253}$.  We assume that the flares decay as

\begin{equation}
$$\[m(t) = m_{0} - 2.5\log \left(\left[\frac{(t-t_{D})}{(t_{0}-t_{D})}\right]^{-5/3}\right) + A_{FUV}\]$$
\end{equation}

\noindent where $(t_{0}-t_{D})=0.11k^{-3/2} (\mbh/10^6 \msun)^{1/2}$; $k =1,2$, or $3$; and $A_{FUV} =8.24\times E(B-V)$ is the Galactic extinction in the FUV \citep{wyd07} measured from the average Milky Way extinction for the field listed in Table 1 of \citet{sch98}, and the \citet{card89} extinction curve.   We record the magnitude of the flaring sources in each epoch of observations for each field with a co-added image with $t_{exp} > 5$ ks, and if the magnitude in an epoch is fainter than the limiting magnitude of the co-added image of the field, it is marked as undetected.  The standard deviation is measured and compared to the 5 $\sigma$ cutoff for each field to see if it would have been selected by our algorithm (described in \S 2).  We then measure the fraction of flares detected to determine $S(\mbh,z)$ for each field.  The error in the detection rate then includes the Poisson error in the number of flares, plus the standard deviation of the fraction events detected between 10 runs of the simulation.

\begin{figure}
\epsscale{.9}
\plotone{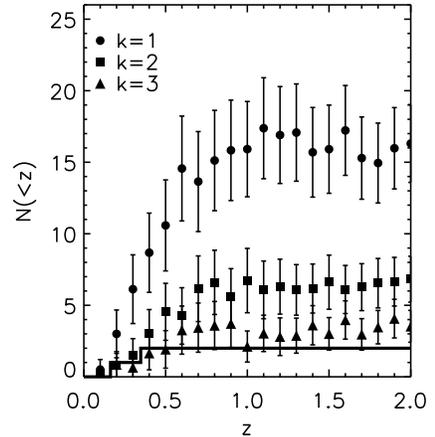}
\caption{Cumulative number of events expected to be detected in our study within a volume out to redshift $z$, for different values of the star's spin up parameter $k$.   The observed cumulative number of events detected is shown with a thick solid line. \label{fig:ntot}}
\end{figure}

In Figure \ref{fig:ntot} we plot the cumulative number of events that would be detected in our search within a redshift volume for different values of the star's spin-up parameter, $k$, compared to our observed cumulative detection of two events within $z < 0.4$. Each redshift volume is from a separate simulation, so it is encouraging that the cumulative distribution converges and is smoothly varying.  The $k=1$ curve, corresponding to a star with no spin, is strongly ruled out because the slow decay of the flares, ($t_{0}-t_{D} \propto k^{-3/2}$), would have resulted in an order of magnitude more flare detections. The $k$=3 curve is in good agreement with our detection rate for $z<0.4$, and the $k=2$ curve is within 1 $\sigma$, but they both over predict the number of events beyond $z\sim0.4$.  
The detection rate for all values of $k$ levels off after $z\sim0.7$, with a total number of detections of $17 \pm 4$, $6 \pm 2$, and $3 \pm 1$ for $k$=1, 2, and 3, respectively.   

\begin{figure}
\epsscale{.9}
\plotone{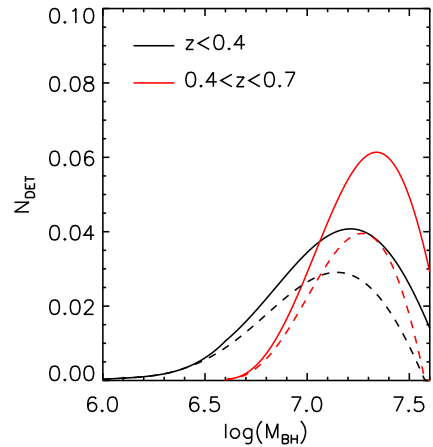}
\caption{{\it Solid lines}: Number of events expected to be detected within a volume $z<0.4$, and $0.4<z<0.7$, as a function of black hole mass, for $k=3$ and the fraction of flares that radiate close to $L_{\rm Edd}$ for 1 $\msun$ stars from Ulmer (1999).   The peak sensitivity of our search shifts from $\mbh=1.6 \times 10^{7} \msun$ for $z<0.4$ to $\mbh=2.2 \times 10^{7} \msun$ for $0.4<z<0.7$, with no sensitivity for $\mbh < 4 \times 10^{6} \msun$ black holes beyond z=0.4.  {\it Dashed lines}: Same for the fraction
of flares that radiate close to $L_{\rm Edd}$ calculated for a more realistic Salpeter mass function from Ulmer (1999), which causes a decrease in number of events expected to be detected in the volume $0.4<z<0.7$.  \label{fig:ntot_comp}}
\end{figure}

The theoretical curves predict the detection of at least one more flare with $0.4 \aplt z \aplt 0.7$.  
Figure \ref{fig:ntot_comp} shows the number of events detected as a function of black hole mass (fitted with a polynomial function for clarity) for the volume $z<0.4$ in comparison to $0.4<z<0.7$.  The peak sensitivity of our search shifts from $\mbh=1.6 \times 10^{7} \msun$ for $z<0.4$ to $\mbh=2.2 \times 10^{7} \msun$ for $0.4<z<0.7$, with no sensitivity to black holes with $\mbh < 4 \times 10^{6} \msun$ beyond $z=0.4$.  Thus, the deficit of detections in this redshift range may be the result of overestimating the fraction of flares from massive black holes that radiate at the Eddington limit.  It is not surprising that the two tidal disruption flares we detected in our search have properties indicating black hole masses of a few $\times 10^{7} \msun$, since that is where our detection sensitivity is at its peak.

We repeated our simulation with a fraction of tidal disruption flares that radiate at the Eddington luminosity calculated for a more realistic Salpeter mass function between 0.3 and 1.0 $\msun$ by \citet{ulmer99}, which gives a lower fraction for high-mass black holes than the curve calculated for 1$\msun$ stars only (equation 8).  We now use $f=1$ for $\mbh < 2.5 \times 10^{6} \msun$, and for $\mbh \ge 2.5 \times 10^{6} \msun$ we use the log-linear function,

\begin{equation}
$$\[f(\mbh) = -0.850 \log(\mbh/\msun)+6.44.\]$$
\end{equation}

\begin{figure}
\epsscale{.9}
\plotone{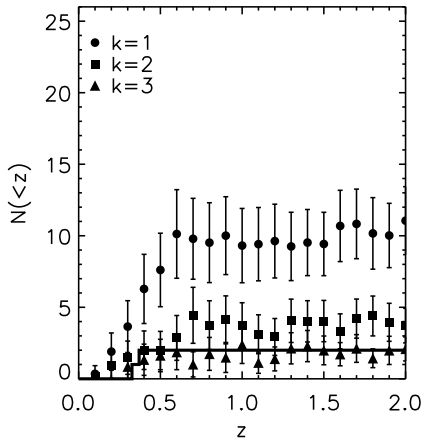}
\caption{Cumulative number of events expected to be detected in our study within a volume out to redshift $z$ for different values of the star's spin up parameter $k$, and the fraction of flares with an Eddington luminosity flare for stars with a Salpeter mass function between 0.3 and 1$\msun$ from Ulmer (1999).  The observed cumulative number of events detected is shown with a thick solid line.\label{fig:salp}}
\end{figure}

\noindent Figure \ref{fig:salp} shows the resulting cumulative detection rate within a redshift volume, with a total number of detections expected of $10\pm3$, $4\pm2$, and $2\pm1$ for $k$=1, 2, and 3, respectively.  The $k=2$ and $k=3$ curves are now both consistent with the observations within 1 $\sigma$, while the $k=1$ curve is still strongly ruled out.  

It is surprising that the models are in such good agreement with our detection rate, since we have not taken into account the effects of extinction internal to the host galaxy, or the inclination angle of the galaxy.  Extinction intrinsic to the host galaxy should suppress the detection rate equally over the range of black hole masses, and thus could allow for lower values of $k$, $k \aplt 2$, to be consistent within 1$\sigma$ of our observed detection rate.  This is important, since the analysis of the individual detections favor smaller values of $k$.

It appears that the values for $\dot{N}(\mbh)$ calculated from the galaxy dynamical models are in good agreement with our detection rates, and can be used along with the observed properties of the flares, to make detailed predictions for the detection capabilities of other surveys.  

\section{Discussion \label{sec:disc}}
For the first time we have measured the broadband SED and detailed light curves of two tidal disruption flare candidates.  The blackbody temperatures, luminosities, and power-law decays of the flares are in excellent agreement with that expected from the simplest ``fallback'' models for the accretion of a tidally disrupted star.  

The statistics of the detection rate of our study require for a star to be spun up on disruption ($k\apgt2$) and for the fraction of flares that radiate at the Eddington luminosity to decline for $\mbh > 2.5 \times 10^{6} \msun$.  However, there is some degeneracy between the effects of internal extinction and the spin-up parameter of the star, and so smaller values of $k$ are compatible with our observations, if we have underestimated the effects of extinction and inclination angle of the host galaxy.  The black hole mass-dependent tidal disruption rates calculated from galaxy dynamical models appear to reproduce the observed rates, given our basic assumptions for the luminosity and temperature of the events.  

The detection of tidal disruption flares in the optical for the first time has important consequences for the detection capabilities of future large optical synoptic surveys such as Pan-STARRS and LSST.  Since the theoretical models for $k=3$ and $f(\mbh)$ for a Salpeter mass function reproduce the observations so well, we can apply our analysis to predict the detection rate of the upcoming Pan-STARRS Medium Deep Survey (MDS), which will monitor an area of $\approx$ 50 deg$^2$ daily with one of five filters ($g,r,i,z$,and $Y$), and with the same filter every 4 days, with a limiting magnitude in the $g$ band of  $m_{g} < 24.76$.  We measure the selection function by simulating flares and requiring that the flare be above the detection threshold of the individual exposures for at least 8 days, so that the power-law decays of a tidal disruption flare can be easily distinguished from optical variability from AGNs.  We do not correct for Galactic extinction, since the MDS fields are at high Galactic latitudes.  

\begin{figure}
\epsscale{.9}
\plotone{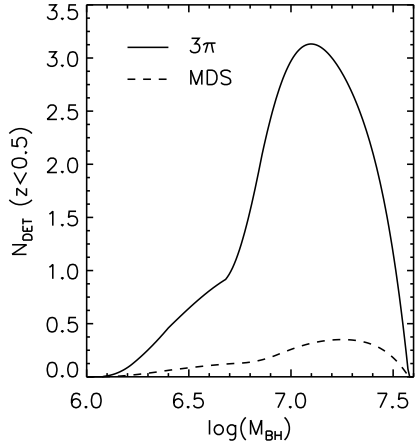}
\caption{Number of events predicted to be detected in the $g$ band per year within a volume out to redshift $z=0.5$ for $k=3$ as a function of black hole mass in the Pan-STARRS MDS and the 3$\pi$ Survey. \label{fig:ntot_opt}}
\end{figure}

Figure \ref{fig:app} shows the predicted peak magnitude of flares in the $g$ band as a function of redshift and black hole mass, and Figure \ref{fig:ntot_opt} shows the detection rate by MDS as a function of black hole mass for $k=3$ and $f(\mbh)$ for a Salpeter mass function, in a volume of $z<0.5$, resulting in a total detection of $\sim$ 15 events yr$^{-1}$.  Figure \ref{fig:ntot_opt} also shows the detection rate of the Pan-STARRS 3$\pi$ survey, which will cover 30,000 deg$^{2}$, with two visits per year with $m_{g} < 23.54$, over 3 yr, resulting in a detection rate of $\sim$130 yr$^{-1}$.  The results of the simulation were fitted with a polynomial for clarity.   The LSST 20,000 deg$^2$ survey will increase these rates by 2 orders of magnitude, since it has a similar depth and cadence to the MDS, with individual visits with $m_{g} < 24.5$, but the factor of 400 increase in survey area that will yield 6000 events yr$^{-1}$!  

The next generation of wide-field optical synoptic surveys will increase the number of tidal disruption flare detections by orders of magnitude, allowing for the study of the ensemble properties and rates of tidal disruption events as a function of galaxy type.  The rates of stellar disruption events are sensitive to the detailed structure of the galaxy nuclei, and can be used to test dynamical models for galaxies that are too distant for detailed spatially resolved observations.  The light curves of individual tidal disruption events can be used as a direct probe of the central black hole masses, and a means to map the dormant black hole mass function in normal galaxies.  The SEDs of events may also be able to probe the spin of black holes, since the characteristic radius of the emission can be compared to the smallest stable particle orbit, which is directly dependent on the spin.
Although the detection sensitivities of these surveys are ``tuned'' to black hole masses of a limited mass range, $(1-3) \times 10^{7} \msun$, they will be a sensitive measure for deviations from the locally established scaling relations between the mass of the black hole and their host galaxy properties.  Evidence for such deviations has already been seen for a sample of 14 Seyfert galaxies at $z\sim0.37$ with $\mbh = 10^{8} - 10^{9} \msun$ by \citet{woo06}.  Large samples of tidal disruption event detections will yield a fruitful data set for testing models for black hole growth and galaxy evolution.

\acknowledgements

We thank our anonymous referee for their insightful comments that helped us improve our paper.
We are grateful for the public database of SN candidates produced by the Supernova Legacy Survey, which has been very useful for this study.  S.G. was supported in part by the Volontariat International-CNES of France, and through \textsl{Chandra} grant G06-7099X
issued by the \textsl{Chandra X-Ray Observatory}, 
which is operated by the Smithsonian Astrophysical Observatory
for and on behalf of NASA.
We gratefully acknowledge NASA's support for construction, 
operation, and science analysis for the \textsl{GALEX} mission, 
developed in cooperation with Centre National d'Etudes Spatiales of France and the Korean Ministry of Science and Technology.
Based on observations obtained with MegaPrime/MegaCam, a joint
project of CFHT and CEA/DAPNIA, at the
Canada-France-Hawaii Telescope (CFHT) which is operated by the National
Research Council (NRC) of
Canada, the Institut National de Sciences de l'Univers of the
Centre National de la Recherche Scientifique (CNRS) of France, and the
University of Hawaii. This work is based in part on data products
produced at TERAPIX and the Canadian Astronomy Data Centre as part of
the CFHT Legacy Survey, a collaborative
project of NRC and CNRS.  This paper makes use of photometric redshifts
produced jointly by Terapix and the VVDS teams, and the CENCOS
interface (http://cencosw.oamp.fr) was used for data retrieval and
analyses.

\clearpage

\input{tab1.tex}

\clearpage
\begin{landscape}
\input{tab2.tex}
\clearpage
\end{landscape}

\clearpage
\begin{landscape}
\input{tab3.tex}
\clearpage
\end{landscape}

\clearpage
\begin{landscape}
\input{tab4.tex}
\clearpage
\end{landscape}

\clearpage
\begin{landscape}
\input{tab5.tex}
\clearpage
\end{landscape}

\clearpage
\input{tab6.tex}
\clearpage

\begin{landscape}
\input{tab7.tex}
\clearpage
\end{landscape}

\clearpage
\begin{landscape}
\input{tab8.tex}
\clearpage
\end{landscape}

\clearpage
\input{tab9.tex}

\input{tab10.tex}

\newpage
\clearpage
\begin{landscape}
\input{tab11.tex}
\clearpage
\end{landscape}

\end{document}

%% file: tab1.tex
\begin{deluxetable}{lrrccrrrrr}
\tablewidth{0pt}
\tablecaption{\textsl{GALEX} Yearly Coadds \label{tab:texp}}
\tablehead{
\colhead{Field} & \colhead{$\alpha$ (J2000)} & \colhead{$\delta$ (J2000)} & \colhead{E(B-V)} & \colhead{Band} & \multicolumn{5}{c}{Exposure Time (ks)} \\
\colhead{} & \colhead{} & \colhead{} & \colhead{} & \colhead{} & \colhead{2003} & \colhead{2004} & \colhead{2005} & \colhead{2006} & \colhead{2007}
}
\startdata
XMM/LSS & 36.3610 & -4.4800 & 0.027 & NUV & 17.8 & 1.1 & 27.5 & 20.0 & \nodata\\
 & & &  & FUV & 17.8 & 1.1 & 22.1 & 20.0 & \nodata\\
\hline
COSMOS & 149.8690 & 2.4558 & 0.018 & NUV & \nodata & 47.5 & \nodata & 8.6 & 20.0\\
 & & &  & FUV & \nodata & 41.5 & \nodata & 8.6 & 20.0\\
\hline
GROTH & 214.9918 & 52.7817 & 0.010 & NUV & 16.1 & 1.5 & 102.4 & 60.1 & 15.7\\
 & & & & FUV & 16.1 & 1.5 & 40.7 & 42.7 & 15.7\\
\hline
CFHTLS D4 & 333.7830 & -17.9350 & 0.027 & NUV & 24.3 & 5.0 & 27.6 & 18.2 & \nodata\\
 & & & & FUV & 24.3 & 5.0 & 5.6 & 18.1 & \nodata
\enddata
\end{deluxetable}

%% file: tab2.tex
\begin{deluxetable}{llcccccccrlccccc}
\tablewidth{0pt}
\tabletypesize{\tiny}
\tablecaption{UV Variable Sources with Optical Matches in XMM/LSS (D1)\label{xmm_tab1}}
\tablehead{
\colhead{No.} & \colhead{\textsl{GALEX} ID} & \colhead{$\alpha$(J2000)} & \colhead{$\delta$(J2000)} & \multicolumn{5}{c}{$FUV$} & \colhead{$\sigma$\tablenotemark{a}} & \colhead{} & \multicolumn{5}{c}{$NUV$} \\
\colhead{} & \colhead{} & \colhead{} & \colhead{} & \colhead{2003} & \colhead{2004} & \colhead{2005} & \colhead{2006} & \colhead{2007} & \colhead{} & \colhead{} & \colhead{2003} & \colhead{2004} & \colhead{2005} & \colhead{2006} & \colhead{2007}
}
\startdata
D1-1 & J022409.7-044007 & 36.0406 & -4.6687 & 23.72$\pm$0.13 & $>$22.8 & 
22.61$\pm$0.05 & 22.93$\pm$0.06 & \nodata &  8.8 & F & 22.72$\pm$0.05 & $>$22.7
 & 22.01$\pm$0.03 & 22.22$\pm$0.03 & \nodata\\
D1-2 & J022409.7-043405 & 36.0406 & -4.5682 & 22.61$\pm$0.06 & 23.00$\pm$0.31 & 
24.00$\pm$0.13 & 23.18$\pm$0.08 & \nodata &  9.4 &     & 21.53$\pm$0.02 & 
22.24$\pm$0.15 & 22.44$\pm$0.04 & 21.98$\pm$0.03 & \nodata\\
D1-3 & J022424.2-043230 & 36.1009 & -4.5418 & 22.44$\pm$0.05 & 23.22$\pm$0.35 & 
23.52$\pm$0.09 & 23.23$\pm$0.08 & \nodata &  8.1 &     & 20.03$\pm$0.01 & 
20.89$\pm$0.06 & 20.98$\pm$0.01 & 21.00$\pm$0.01 & \nodata\\
D1-4 & J022434.7-045350 & 36.1444 & -4.8973 & $>$24.3 & $>$22.8 & 23.91$\pm$0.12
 & $>$24.3 & \nodata &  6.0 & T & $>$24.2 & $>$22.7 & 20.38$\pm$0.01 & $>$24.3
 & \nodata\\
D1-5 & J022449.2-041802 & 36.2052 & -4.3006 & 22.79$\pm$0.07 & $>$22.8 & 
24.04$\pm$0.13 & 24.80$\pm$0.24 & \nodata & 16.1 & F & 22.12$\pm$0.03 & $>$22.7
 & 22.75$\pm$0.05 & 23.16$\pm$0.07 & \nodata\\
D1-6 & J022449.9-043025 & 36.2078 & -4.5072 & 22.92$\pm$0.07 & 22.34$\pm$0.20 & 
23.54$\pm$0.09 & 23.89$\pm$0.12 & \nodata &  6.3 &     & 21.42$\pm$0.02 & 
20.93$\pm$0.06 & 21.71$\pm$0.02 & 21.93$\pm$0.03 & \nodata\\
D1-7 & J022451.2-040527 & 36.2132 & -4.0911 & 22.47$\pm$0.06 & 23.16$\pm$0.35 & 
22.55$\pm$0.05 & 23.16$\pm$0.08 & \nodata &  5.2 &     & 21.27$\pm$0.02 & 
21.50$\pm$0.09 & 21.34$\pm$0.02 & 21.78$\pm$0.02 & \nodata\\
D1-8 & J022503.2-040539 & 36.2635 & -4.0944 & 22.77$\pm$0.07 & 23.35$\pm$0.39 & 
21.43$\pm$0.03 & 21.40$\pm$0.03 & \nodata & 14.7 &     & 21.46$\pm$0.02 & 
22.20$\pm$0.14 & 20.65$\pm$0.01 & 20.58$\pm$0.01 & \nodata\\
D1-9 & J022517.0-043258 & 36.3207 & -4.5497 & $>$24.3 & 23.23$\pm$0.36 & 
23.41$\pm$0.08 & 24.41$\pm$0.18 & \nodata &  7.7 & T & $>$24.2 & 23.05$\pm$0.27
 & 23.32$\pm$0.07 & 24.00$\pm$0.13 & \nodata\\
D1-10 & J022518.1-043155 & 36.3254 & -4.5320 & 23.86$\pm$0.14 & $>$22.8 & 
$>$24.4 & $>$24.3 & \nodata &  6.9 & F & 23.47$\pm$0.10 & $>$22.7 & $>$24.4 & 
$>$24.3 & \nodata\\
D1-11 & J022526.5-040305 & 36.3604 & -4.0515 & $>$24.3 & $>$22.8 & 
24.38$\pm$0.17 & 23.65$\pm$0.11 & \nodata &  5.5 & F & $>$24.2 & $>$22.7 & 
22.79$\pm$0.05 & 22.06$\pm$0.03 & \nodata\\
D1-12 & J022527.5-035954 & 36.3647 & -3.9984 & 23.89$\pm$0.15 & 23.17$\pm$0.35
 & 23.64$\pm$0.10 & 24.72$\pm$0.23 & \nodata &  5.7 &     & 22.69$\pm$0.05 & 
22.09$\pm$0.13 & 22.82$\pm$0.05 & 23.75$\pm$0.11 & \nodata\\
D1-13 & J022539.4-042228 & 36.4141 & -4.3746 & 22.12$\pm$0.04 & 23.23$\pm$0.36
 & 22.45$\pm$0.05 & 24.55$\pm$0.20 & \nodata & 25.0 &     & 21.25$\pm$0.02 & 
21.97$\pm$0.12 & 21.76$\pm$0.02 & 22.86$\pm$0.05 & \nodata\\
D1-14 & J022544.4-040102 & 36.4350 & -4.0173 & 22.65$\pm$0.06 & 23.28$\pm$0.38
 & 23.18$\pm$0.07 & 23.93$\pm$0.13 & \nodata &  8.6 & F & 21.85$\pm$0.03 & 
21.81$\pm$0.11 & 22.12$\pm$0.03 & 22.54$\pm$0.04 & \nodata\\
D1-15 & J022552.2-042249 & 36.4674 & -4.3803 & 22.56$\pm$0.06 & 22.12$\pm$0.18
 & 22.76$\pm$0.05 & 23.55$\pm$0.10 & \nodata &  7.4 &     & 21.31$\pm$0.02 & 
21.05$\pm$0.07 & 21.50$\pm$0.02 & 21.91$\pm$0.03 & \nodata\\
D1-16 & J022555.4-043918 & 36.4810 & -4.6551 & 21.96$\pm$0.04 & 22.36$\pm$0.21
 & 22.26$\pm$0.04 & 22.73$\pm$0.06 & \nodata &  6.3 & F & 21.26$\pm$0.02 & 
21.32$\pm$0.08 & 21.40$\pm$0.02 & 21.71$\pm$0.02 & \nodata\\
D1-17 & J022556.9-045853 & 36.4869 & -4.9815 & 23.09$\pm$0.08 & 22.65$\pm$0.25
 & 21.85$\pm$0.03 & 22.77$\pm$0.06 & \nodata & 12.7 &     & 21.88$\pm$0.03 & 
21.41$\pm$0.08 & 21.09$\pm$0.01 & 21.52$\pm$0.02 & \nodata\\
D1-18 & J022558.1-045720 & 36.4920 & -4.9558 & 22.88$\pm$0.07 & $>$22.8 & 
21.99$\pm$0.04 & 22.97$\pm$0.07 & \nodata & 10.4 & F & 21.11$\pm$0.02 & $>$22.7
 & 20.68$\pm$0.01 & 21.17$\pm$0.02 & \nodata\\
D1-19 & J022606.7-045723 & 36.5281 & -4.9565 & 21.48$\pm$0.03 & 21.34$\pm$0.12
 & 22.27$\pm$0.04 & 21.58$\pm$0.03 & \nodata &  7.3 &     & 20.89$\pm$0.02 & 
20.77$\pm$0.06 & 21.41$\pm$0.02 & 20.82$\pm$0.01 & \nodata\\
D1-20 & J022639.8-042004 & 36.6660 & -4.3345 & 22.61$\pm$0.06 & $>$22.8 & 
23.23$\pm$0.07 & 23.55$\pm$0.10 & \nodata &  6.1 & F & 21.87$\pm$0.03 & $>$22.7
 & 22.35$\pm$0.03 & 22.68$\pm$0.04 & \nodata\\
D1-21 & J022647.8-043135 & 36.6993 & -4.5265 & 23.64$\pm$0.12 & $>$22.8 & 
$>$24.4 & $>$24.3 & \nodata &  8.8 & F & 22.49$\pm$0.04 & $>$22.7 & $>$24.4 & 
$>$24.3 & \nodata\\
D1-22 & J022701.5-040913 & 36.7562 & -4.1537 & 23.07$\pm$0.08 & $>$22.8 & 
24.04$\pm$0.13 & 23.65$\pm$0.10 & \nodata &  6.2 &     & 21.26$\pm$0.02 & 
$>$22.7 & 21.92$\pm$0.03 & 21.74$\pm$0.02 & \nodata\\
D1-23 & J022716.1-044539 & 36.8172 & -4.7609 & 22.01$\pm$0.04 & 21.98$\pm$0.17
 & 21.62$\pm$0.03 & 22.69$\pm$0.06 & \nodata &  9.9 &     & 20.69$\pm$0.01 & 
20.63$\pm$0.05 & 20.56$\pm$0.01 & 21.09$\pm$0.02 & \nodata\\
\enddata
\tablenotetext{a}{FUV flares that fade monotonically, have a steady flux before the flare, and do not fade below the steady-state flux are labeled with an 'F'. Transient FUV flares that fade monotonically, and are not detected before the start of the flare are labeled with a 'T'.}
\end{deluxetable}

%% file: tab3.tex
\begin{deluxetable}{llcccccccrlccccc}
\tablewidth{0pt}
\tabletypesize{\tiny}
\tablecaption{UV Variable Sources with Optical Matches in COSMOS (D2)\label{cosmos_03_tab1}}
\tablehead{
\colhead{No.} & \colhead{\textsl{GALEX} ID} & \colhead{$\alpha$(J2000)} & \colhead{$\delta$(J2000)} & \multicolumn{5}{c}{$FUV$} & \colhead{$\sigma$\tablenotemark{a}} & \colhead{} & \multicolumn{5}{c}{$NUV$} \\
\colhead{} & \colhead{} & \colhead{} & \colhead{} & \colhead{2003} & \colhead{2004} & \colhead{2005} & \colhead{2006} & \colhead{2007} & \colhead{} & \colhead{} & \colhead{2003} & \colhead{2004} & \colhead{2005} & \colhead{2006} & \colhead{2007}
}
\startdata
D2-1 & J095829.3+021542 & 149.6219 & 2.2619 & \nodata & 22.42$\pm$0.03 & \nodata
 & 22.12$\pm$0.06 & 22.78$\pm$0.06 &  6.3 &     & \nodata & 21.69$\pm$0.02 & 
\nodata & 21.51$\pm$0.03 & 21.85$\pm$0.02\\
D2-2 & J095848.9+023441 & 149.7037 & 2.5782 & \nodata & 23.67$\pm$0.07 & \nodata
 & 25.21$\pm$0.51 & $>$24.3 &  6.7 & F & \nodata & 21.27$\pm$0.01 & \nodata & 
22.13$\pm$0.05 & $>$24.2\\
D2-3 & J095858.6+021459 & 149.7440 & 2.2498 & \nodata & 20.66$\pm$0.01 & \nodata
 & 21.43$\pm$0.04 & 21.46$\pm$0.03 & 17.1 & F & \nodata & 20.61$\pm$0.01 & 
\nodata & 21.29$\pm$0.03 & 21.14$\pm$0.02\\
D2-4 & J095921.5+024029 & 149.8394 & 2.6748 & \nodata & 21.15$\pm$0.02 & \nodata
 & 21.34$\pm$0.04 & 21.57$\pm$0.03 &  6.9 & F & \nodata & 20.38$\pm$0.01 & 
\nodata & 20.56$\pm$0.02 & 20.76$\pm$0.01\\
D2-5 & J095947.0+022209 & 149.9457 & 2.3693 & \nodata & 22.43$\pm$0.03 & \nodata
 & 22.18$\pm$0.06 & 23.19$\pm$0.07 &  7.6 &     & \nodata & 21.36$\pm$0.01 & 
\nodata & 21.22$\pm$0.03 & 22.03$\pm$0.03\\
D2-6 & J095954.8+021706 & 149.9784 & 2.2852 & \nodata & 24.13$\pm$0.10 & \nodata
 & 22.90$\pm$0.10 & 24.39$\pm$0.16 &  6.9 & F & \nodata & 22.02$\pm$0.02 & 
\nodata & 21.40$\pm$0.03 & 22.03$\pm$0.03\\
D2-7 & J095958.5+021530 & 149.9938 & 2.2585 & \nodata & 22.95$\pm$0.04 & \nodata
 & $>$24.2 & $>$24.3 & 16.7 & F & \nodata & 21.62$\pm$0.02 & \nodata & $>$24.1
 & $>$24.2\\
D2-8 & J095958.9+020023 & 149.9955 & 2.0065 & \nodata & 25.40$\pm$0.27 & \nodata
 & 23.15$\pm$0.11 & 23.61$\pm$0.10 & 13.2 & F & \nodata & 23.79$\pm$0.08 & 
\nodata & 22.28$\pm$0.05 & 22.38$\pm$0.04\\
D2-9 & J100002.0+024216 & 150.0085 & 2.7045 & \nodata & 22.77$\pm$0.04 & \nodata
 & 23.89$\pm$0.19 & 24.57$\pm$0.19 & 19.6 & F & \nodata & 22.24$\pm$0.02 & 
\nodata & 22.84$\pm$0.08 & 23.18$\pm$0.07\\
D2-10 & J100012.9+023522 & 150.0539 & 2.5897 & \nodata & 20.57$\pm$0.01 & 
\nodata & 20.89$\pm$0.03 & 20.63$\pm$0.02 &  5.6 &     & \nodata & 
19.26$\pm$0.00 & \nodata & 19.46$\pm$0.01 & 19.23$\pm$0.01\\
D2-11 & J100014.9+022717 & 150.0622 & 2.4550 & \nodata & 22.67$\pm$0.04 & 
\nodata & 23.05$\pm$0.10 & 23.33$\pm$0.08 &  5.9 & F & \nodata & 21.53$\pm$0.01
 & \nodata & 21.72$\pm$0.04 & 21.92$\pm$0.03\\
D2-12 & J100017.6+020011 & 150.0732 & 2.0032 & \nodata & 22.82$\pm$0.04 & 
\nodata & 22.91$\pm$0.10 & 23.82$\pm$0.11 &  8.9 & F & \nodata & 22.32$\pm$0.02
 & \nodata & 22.42$\pm$0.06 & 23.74$\pm$0.10\\
D2-13 & J100024.7+023149 & 150.1028 & 2.5304 & \nodata & 23.50$\pm$0.06 & 
\nodata & 24.05$\pm$0.21 & 24.50$\pm$0.18 &  5.7 & F & \nodata & 19.18$\pm$0.00
 & \nodata & 19.87$\pm$0.01 & 19.81$\pm$0.01\\
D2-14 & J100029.7+022129 & 150.1237 & 2.3583 & \nodata & 23.89$\pm$0.08 & 
\nodata & 23.54$\pm$0.14 & 22.98$\pm$0.06 &  5.4 &     & \nodata & 
22.19$\pm$0.02 & \nodata & 22.11$\pm$0.05 & 21.82$\pm$0.02\\
D2-15 & J100035.0+020235 & 150.1458 & 2.0431 & \nodata & 22.41$\pm$0.03 & 
\nodata & 22.17$\pm$0.06 & 23.52$\pm$0.09 & 11.0 &     & \nodata & 
21.14$\pm$0.01 & \nodata & 21.14$\pm$0.02 & 21.91$\pm$0.03\\
D2-16 & J100046.8+020404 & 150.1949 & 2.0680 & \nodata & 21.47$\pm$0.02 & 
\nodata & 21.95$\pm$0.06 & 21.86$\pm$0.03 &  7.6 & F & \nodata & 20.80$\pm$0.01
 & \nodata & 21.16$\pm$0.02 & 21.18$\pm$0.02\\
D2-17 & J100055.4+023441 & 150.2309 & 2.5781 & \nodata & 22.15$\pm$0.03 & 
\nodata & 23.66$\pm$0.16 & 22.64$\pm$0.05 & 19.2 &     & \nodata & 
21.04$\pm$0.01 & \nodata & 22.34$\pm$0.05 & 21.52$\pm$0.02\\
D2-18 & J100104.2+023349 & 150.2677 & 2.5637 & \nodata & $>$25.0 & \nodata & 
$>$24.2 & 23.28$\pm$0.08 & 15.7 & T & \nodata & $>$25.0 & \nodata & $>$24.1 & 
$>$24.2\\
D2-19 & J100104.3+023402 & 150.2678 & 2.5673 & \nodata & 23.42$\pm$0.06 & 
\nodata & 23.74$\pm$0.17 & 22.39$\pm$0.05 & 12.0 &     & \nodata & 
22.23$\pm$0.02 & \nodata & 22.37$\pm$0.06 & 21.50$\pm$0.02\\
D2-20 & J100104.8+023329 & 150.2699 & 2.5583 & \nodata & $>$25.0 & \nodata & 
$>$24.2 & 23.76$\pm$0.11 &  8.1 & T & \nodata & $>$25.0 & \nodata & $>$24.1 & 
$>$24.2\\
D2-21 & J100105.4+023323 & 150.2723 & 2.5564 & \nodata & 23.71$\pm$0.08 & 
\nodata & 23.66$\pm$0.16 & 22.72$\pm$0.06 &  7.1 & F & \nodata & 23.02$\pm$0.04
 & \nodata & 22.86$\pm$0.08 & 21.97$\pm$0.03\\
D2-22 & J100116.3+023607 & 150.3180 & 2.6021 & \nodata & 22.22$\pm$0.03 & 
\nodata & 23.54$\pm$0.15 & 22.71$\pm$0.06 & 16.6 &     & \nodata & 
21.15$\pm$0.01 & \nodata & 21.84$\pm$0.04 & 21.36$\pm$0.02\\
D2-23 & J100124.0+021445 & 150.3500 & 2.2461 & \nodata & 23.39$\pm$0.06 & 
\nodata & 24.12$\pm$0.22 & 23.39$\pm$0.08 &  5.4 &     & \nodata & 
22.29$\pm$0.02 & \nodata & 22.61$\pm$0.07 & 22.24$\pm$0.03\\
\enddata
\tablenotetext{a}{FUV flares that fade monotonically, have a steady flux before the flare, and do not fade below the steady-state flux are labeled with an 'F'. Transient FUV flares that fade monotonically, and are not detected before the start of the flare are labeled with a 'T'.}
\end{deluxetable}

%% file: tab4.tex
\begin{deluxetable}{llcccccccrlccccc}
\tablewidth{0pt}
\tabletypesize{\tiny}
\tablecaption{UV Variable Sources with Optical Matches in GROTH (D3)\label{groth_tab1}}
\tablehead{
\colhead{No.} & \colhead{\textsl{GALEX} ID} & \colhead{$\alpha$(J2000)} & \colhead{$\delta$(J2000)} & \multicolumn{5}{c}{$FUV$} & \colhead{$\sigma$\tablenotemark{a}} & \colhead{} & \multicolumn{5}{c}{$NUV$} \\
\colhead{} & \colhead{} & \colhead{} & \colhead{} & \colhead{2003} & \colhead{2004} & \colhead{2005} & \colhead{2006} & \colhead{2007} & \colhead{} & \colhead{} & \colhead{2003} & \colhead{2004} & \colhead{2005} & \colhead{2006} & \colhead{2007}
}
\startdata
D3-1 & J141714.4+525130 & 214.3099 & 52.8585 & 23.91$\pm$0.16 & $>$23.2 & 
$>$25.0 & $>$25.0 & $>$24.5 &  7.8 & F & 23.46$\pm$0.10 & $>$23.1 & $>$25.4 & 
$>$25.1 & $>$24.4\\
D3-2 & J141715.2+530303 & 214.3135 & 53.0510 & 24.30$\pm$0.22 & $>$23.2 & 
23.57$\pm$0.07 & 25.25$\pm$0.24 & 24.22$\pm$0.18 &  7.1 &     & 23.22$\pm$0.08
 & $>$23.1 & 22.82$\pm$0.03 & 24.06$\pm$0.09 & 23.10$\pm$0.07\\
D3-3 & J141721.2+530916 & 214.3382 & 53.1546 & $>$24.5 & $>$23.2 & 
22.17$\pm$0.03 & 23.43$\pm$0.06 & $>$24.5 & 25.4 & T & $>$24.4 & $>$23.1 & 
22.04$\pm$0.02 & 22.85$\pm$0.03 & $>$24.4\\
D3-4 & J141724.6+523025 & 214.3527 & 52.5071 & 21.27$\pm$0.03 & 21.26$\pm$0.10
 & 20.94$\pm$0.02 & 21.42$\pm$0.02 & 22.04$\pm$0.04 &  8.6 &     & 
20.54$\pm$0.01 & 20.53$\pm$0.04 & 20.28$\pm$0.01 & 20.69$\pm$0.01 & 
21.16$\pm$0.02\\
D3-5 & J141734.9+522811 & 214.3955 & 52.4698 & 24.54$\pm$0.27 & $>$23.2 & 
24.01$\pm$0.10 & 22.51$\pm$0.03 & 23.92$\pm$0.14 & 17.3 &     & 22.86$\pm$0.06
 & $>$23.1 & 22.34$\pm$0.02 & 21.39$\pm$0.01 & 22.28$\pm$0.04\\
D3-6 & J141801.5+525201 & 214.5064 & 52.8670 & 21.47$\pm$0.03 & 21.22$\pm$0.09
 & 21.49$\pm$0.02 & 22.11$\pm$0.03 & 21.85$\pm$0.04 &  5.1 &     & 
20.45$\pm$0.01 & 20.21$\pm$0.03 & 20.41$\pm$0.01 & 20.79$\pm$0.01 & 
20.71$\pm$0.01\\
D3-7 & J141830.3+522213 & 214.6261 & 52.3703 & 21.53$\pm$0.03 & 21.81$\pm$0.13
 & 22.36$\pm$0.03 & 23.69$\pm$0.07 & 24.26$\pm$0.19 & 21.3 & F & 20.64$\pm$0.01
 & 20.95$\pm$0.05 & 21.19$\pm$0.01 & 21.89$\pm$0.02 & 22.44$\pm$0.04\\
D3-8 & J141833.5+530733 & 214.6394 & 53.1259 & $>$24.5 & $>$23.2 & 
23.79$\pm$0.08 & $>$25.0 & 24.79$\pm$0.28 &  7.5 & T & $>$24.4 & $>$23.1 & 
22.78$\pm$0.03 & $>$25.1 & 23.15$\pm$0.07\\
D3-9 & J141834.9+524205 & 214.6453 & 52.7015 & 24.03$\pm$0.18 & $>$23.2 & 
$>$25.0 & $>$25.0 & $>$24.5 &  5.3 & F & 23.75$\pm$0.12 & $>$23.1 & $>$25.4 & 
$>$25.1 & $>$24.4\\
D3-10 & J141903.9+530856 & 214.7663 & 53.1489 & 23.39$\pm$0.11 & 23.38$\pm$0.35
 & $>$25.0 & 24.26$\pm$0.11 & 23.70$\pm$0.12 &  6.3 &     & 22.56$\pm$0.05 & 
22.77$\pm$0.18 & $>$25.4 & 22.76$\pm$0.03 & 22.46$\pm$0.04\\
D3-11 & J141905.2+522528 & 214.7716 & 52.4246 & 21.73$\pm$0.04 & 21.44$\pm$0.11
 & 21.23$\pm$0.02 & 21.15$\pm$0.02 & 21.58$\pm$0.03 &  5.4 &     & 
20.76$\pm$0.01 & 20.39$\pm$0.04 & 20.19$\pm$0.01 & 20.31$\pm$0.01 & 
20.58$\pm$0.01\\
D3-12 & J141918.1+524158 & 214.8253 & 52.6997 & 22.56$\pm$0.06 & 22.50$\pm$0.19
 & 22.06$\pm$0.03 & 22.69$\pm$0.04 & 23.56$\pm$0.11 & 10.4 &     & 
21.87$\pm$0.03 & 21.88$\pm$0.09 & 21.45$\pm$0.01 & 22.00$\pm$0.02 & 
22.75$\pm$0.05\\
D3-13 & J141929.8+525206 & 214.8741 & 52.8684 & $>$24.5 & 23.18$\pm$0.30 & 
23.41$\pm$0.06 & 24.32$\pm$0.11 & 24.34$\pm$0.20 &  5.8 & T & $>$24.4 & 
22.21$\pm$0.12 & 22.67$\pm$0.03 & $>$25.1 & $>$24.4\\
D3-14 & J141936.1+521359 & 214.9006 & 52.2331 & 24.03$\pm$0.18 & $>$23.2 & 
$>$25.0 & $>$25.0 & $>$24.5 &  5.3 & F & 24.17$\pm$0.17 & $>$23.1 & $>$25.4 & 
$>$25.1 & $>$24.4\\
D3-15 & J141946.4+525943 & 214.9434 & 52.9953 & 23.56$\pm$0.12 & 22.93$\pm$0.25
 & 23.08$\pm$0.05 & 22.74$\pm$0.04 & 22.76$\pm$0.07 &  5.2 &     & 
21.84$\pm$0.03 & 21.78$\pm$0.09 & 21.68$\pm$0.02 & 21.43$\pm$0.01 & 
21.29$\pm$0.02\\
D3-16 & J141951.5+524205 & 214.9646 & 52.7015 & 22.68$\pm$0.07 & $>$23.2 & 
23.47$\pm$0.07 & 24.04$\pm$0.09 & 23.15$\pm$0.08 &  5.9 &     & 22.16$\pm$0.04
 & $>$23.1 & 22.66$\pm$0.03 & 23.02$\pm$0.04 & 22.53$\pm$0.04\\
D3-17 & J141959.8+524245 & 214.9991 & 52.7126 & 23.47$\pm$0.11 & 22.85$\pm$0.24
 & 22.79$\pm$0.04 & 22.88$\pm$0.04 & 22.48$\pm$0.06 &  5.3 &     & 
23.11$\pm$0.07 & 22.99$\pm$0.21 & 22.60$\pm$0.03 & 22.64$\pm$0.03 & 
22.32$\pm$0.04\\
D3-18 & J142039.5+521929 & 215.1646 & 52.3248 & 21.50$\pm$0.03 & 21.61$\pm$0.12
 & 21.95$\pm$0.03 & 21.86$\pm$0.02 & 22.31$\pm$0.05 &  5.6 &     & 
20.38$\pm$0.01 & 20.47$\pm$0.04 & 20.64$\pm$0.01 & 20.72$\pm$0.01 & 
21.16$\pm$0.02\\
D3-19 & J142048.2+530816 & 215.2010 & 53.1379 & $>$24.5 & $>$23.2 & 
22.80$\pm$0.04 & 23.65$\pm$0.07 & $>$24.5 & 13.1 & T & $>$24.4 & $>$23.1 & 
22.42$\pm$0.03 & 23.27$\pm$0.05 & $>$24.4\\
D3-20 & J142056.9+524830 & 215.2370 & 52.8084 & 21.83$\pm$0.04 & 22.33$\pm$0.17
 & 22.33$\pm$0.03 & 21.66$\pm$0.02 & 22.34$\pm$0.05 &  6.2 &     & 
20.78$\pm$0.01 & 21.00$\pm$0.05 & 21.08$\pm$0.01 & 20.52$\pm$0.01 & 
21.07$\pm$0.02\\
D3-21 & J142107.7+530319 & 215.2821 & 53.0553 & 23.81$\pm$0.15 & $>$23.2 & 
22.75$\pm$0.04 & 23.33$\pm$0.06 & 24.09$\pm$0.17 &  8.0 & F & 22.40$\pm$0.04 & 
$>$23.1 & 22.02$\pm$0.02 & 22.22$\pm$0.02 & 22.70$\pm$0.05\\
D3-22 & J142112.3+524147 & 215.3012 & 52.6965 & 21.01$\pm$0.03 & 21.61$\pm$0.12
 & 21.67$\pm$0.02 & 21.85$\pm$0.02 & 21.61$\pm$0.03 &  6.0 &     & 
20.30$\pm$0.01 & 20.70$\pm$0.05 & 20.76$\pm$0.01 & 20.77$\pm$0.01 & 
20.64$\pm$0.01\\
D3-23 & J142113.9+521747 & 215.3081 & 52.2966 & 22.50$\pm$0.06 & 22.32$\pm$0.17
 & 22.34$\pm$0.03 & 22.06$\pm$0.03 & 23.06$\pm$0.08 &  7.0 &     & 
22.36$\pm$0.04 & 22.23$\pm$0.12 & 22.23$\pm$0.02 & 21.96$\pm$0.02 & 
22.90$\pm$0.06\\
D3-24 & J142115.7+523156 & 215.3154 & 52.5323 & 23.89$\pm$0.16 & $>$23.2 & 
23.86$\pm$0.09 & 23.14$\pm$0.05 & 24.01$\pm$0.16 &  5.5 &     & 22.47$\pm$0.05
 & $>$23.1 & 22.68$\pm$0.03 & 22.19$\pm$0.02 & 22.96$\pm$0.06\\
D3-25 & J142125.0+521800 & 215.3542 & 52.3001 & $>$24.5 & $>$23.2 & 
25.49$\pm$0.31 & 22.29$\pm$0.03 & 25.09$\pm$0.37 & 33.6 & F & $>$24.4 & $>$23.1
 & 23.04$\pm$0.04 & 21.24$\pm$0.01 & 23.13$\pm$0.07\\
D3-26 & J142133.9+530245 & 215.3912 & 53.0460 & 23.86$\pm$0.16 & 23.51$\pm$0.38
 & 23.35$\pm$0.06 & 23.23$\pm$0.05 & 22.53$\pm$0.06 &  5.9 &     & 
22.39$\pm$0.04 & 22.00$\pm$0.10 & 22.07$\pm$0.02 & 22.04$\pm$0.02 & 
21.61$\pm$0.02\\
D3-27 & J142134.2+523429 & 215.3925 & 52.5749 & 23.75$\pm$0.14 & $>$23.2 & 
25.10$\pm$0.23 & 23.91$\pm$0.08 & 23.75$\pm$0.13 &  5.5 &     & 21.15$\pm$0.02
 & $>$23.1 & 21.67$\pm$0.02 & 21.21$\pm$0.01 & 21.15$\pm$0.02\\
D3-28 & J142135.9+523139 & 215.3996 & 52.5277 & 20.49$\pm$0.02 & 21.09$\pm$0.09
 & 20.07$\pm$0.01 & 20.56$\pm$0.01 & 21.03$\pm$0.03 &  7.5 &     & 
20.28$\pm$0.01 & 20.77$\pm$0.05 & 19.95$\pm$0.01 & 20.17$\pm$0.01 & 
20.63$\pm$0.01\\
D3-29 & J142151.1+524951 & 215.4629 & 52.8309 & 22.24$\pm$0.05 & $>$23.2 & 
21.67$\pm$0.02 & 22.19$\pm$0.03 & 22.21$\pm$0.05 &  5.3 &     & 21.85$\pm$0.03
 & $>$23.1 & 21.44$\pm$0.01 & 21.84$\pm$0.02 & 21.84$\pm$0.03\\
D3-30 & J142206.8+524957 & 215.5285 & 52.8327 & 23.23$\pm$0.10 & $>$23.2 & 
$>$25.0 & 23.66$\pm$0.07 & 24.16$\pm$0.17 &  7.3 &     & 22.68$\pm$0.05 & 
$>$23.1 & $>$25.4 & 22.79$\pm$0.03 & 23.11$\pm$0.07\\
D3-31 & J142209.1+530559 & 215.5381 & 53.1000 & 20.94$\pm$0.02 & 21.11$\pm$0.09
 & 20.88$\pm$0.02 & 21.67$\pm$0.02 & 21.38$\pm$0.03 &  6.4 &     & 
19.81$\pm$0.01 & 19.93$\pm$0.03 & 19.85$\pm$0.01 & 20.26$\pm$0.01 & 
20.15$\pm$0.01\\
D3-32 & J142243.5+530520 & 215.6814 & 53.0890 & 22.39$\pm$0.06 & $>$23.2 & 
22.09$\pm$0.03 & 21.65$\pm$0.02 & 23.21$\pm$0.09 & 13.7 &     & 21.40$\pm$0.02
 & $>$23.1 & 21.06$\pm$0.01 & 20.65$\pm$0.01 & 21.72$\pm$0.03\\
\enddata
\tablenotetext{a}{FUV flares that fade monotonically, have a steady flux before the flare, and do not fade below the steady-state flux are labeled with an 'F'. Transient FUV flares that fade monotonically, and are not detected before the start of the flare are labeled with a 'T'.}
\end{deluxetable}

%% file: tab5.tex
\begin{deluxetable}{llcccccccrlccccc}
\tablewidth{0pt}
\tabletypesize{\tiny}
\tablecaption{UV Variable Sources with Optical Matches in CFHTLS D4\label{d4_tab1}}
\tablehead{
\colhead{No.} & \colhead{\textsl{GALEX} ID} & \colhead{$\alpha$(J2000)} & \colhead{$\delta$(J2000)} & \multicolumn{5}{c}{$FUV$} & \colhead{$\sigma$\tablenotemark{a}} & \colhead{} & \multicolumn{5}{c}{$NUV$} \\
\colhead{} & \colhead{} & \colhead{} & \colhead{} & \colhead{2003} & \colhead{2004} & \colhead{2005} & \colhead{2006} & \colhead{2007} & \colhead{} & \colhead{} & \colhead{2003} & \colhead{2004} & \colhead{2005} & \colhead{2006} & \colhead{2007}
}
\startdata
D4-1 & J221336.1-180531 & 333.4003 & -18.0920 & 21.25$\pm$0.02 & 21.15$\pm$0.05
 & 22.28$\pm$0.08 & 21.34$\pm$0.03 & \nodata & 12.7 &     & 20.20$\pm$0.01 & 
20.07$\pm$0.02 & 20.97$\pm$0.03 & 20.41$\pm$0.01 & \nodata\\
D4-2 & J221409.7-175652 & 333.5403 & -17.9480 & 21.25$\pm$0.02 & 21.15$\pm$0.05
 & 21.06$\pm$0.04 & 20.72$\pm$0.02 & \nodata &  5.4 &     & 20.31$\pm$0.01 & 
20.31$\pm$0.02 & 20.13$\pm$0.02 & 19.94$\pm$0.01 & \nodata\\
D4-3 & J221413.0-172909 & 333.5542 & -17.4860 & 22.66$\pm$0.05 & 22.41$\pm$0.10
 & 21.65$\pm$0.06 & 23.16$\pm$0.08 & \nodata &  8.7 &     & 21.18$\pm$0.02 & 
20.90$\pm$0.03 & 20.33$\pm$0.02 & 21.69$\pm$0.02 & \nodata\\
D4-4 & J221429.5-173007 & 333.6230 & -17.5020 & 19.29$\pm$0.01 & 19.28$\pm$0.02
 & 19.71$\pm$0.02 & 19.66$\pm$0.01 & \nodata &  5.4 &     & 18.37$\pm$0.00 & 
18.35$\pm$0.01 & 18.65$\pm$0.01 & 18.65$\pm$0.00 & \nodata\\
D4-5 & J221432.7-180158 & 333.6363 & -18.0330 & 21.41$\pm$0.03 & 21.80$\pm$0.07
 & 21.11$\pm$0.04 & 22.01$\pm$0.04 & \nodata &  7.2 &     & 20.19$\pm$0.01 & 
20.51$\pm$0.02 & 19.84$\pm$0.01 & 20.40$\pm$0.01 & \nodata\\
D4-6 & J221433.8-173007 & 333.6408 & -17.5020 & 22.91$\pm$0.06 & 23.55$\pm$0.20
 & 23.26$\pm$0.15 & 22.57$\pm$0.05 & \nodata &  5.5 &     & 22.28$\pm$0.03 & 
22.49$\pm$0.08 & 22.38$\pm$0.07 & 22.00$\pm$0.03 & \nodata\\
D4-7 & J221508.9-175602 & 333.7869 & -17.9340 & 23.43$\pm$0.08 & $>$23.6 & 
$>$23.7 & $>$24.3 & \nodata &  6.5 & F & 22.50$\pm$0.04 & $>$23.5 & $>$24.4 & 
$>$24.2 & \nodata\\
D4-8 & J221639.6-174427 & 334.1651 & -17.7410 & 22.83$\pm$0.06 & 21.28$\pm$0.05
 & 21.88$\pm$0.07 & 21.54$\pm$0.03 & \nodata & 12.2 &     & 21.37$\pm$0.02 & 
20.41$\pm$0.02 & 20.96$\pm$0.03 & 20.56$\pm$0.01 & \nodata\\
D4-9 & J221647.8-175454 & 334.1991 & -17.9150 & 22.41$\pm$0.04 & 21.91$\pm$0.07
 & 22.35$\pm$0.09 & 22.82$\pm$0.06 & \nodata &  5.2 &     & 21.23$\pm$0.02 & 
20.90$\pm$0.03 & 21.19$\pm$0.03 & 21.53$\pm$0.02 & \nodata\\
\enddata
\tablenotetext{a}{FUV flares that fade monotonically, have a steady flux before the flare, and do not fade below the steady-state flux are labeled with an 'F'. Transient FUV flares that fade monotonically, and are not detected before the start of the flare are labeled with a 'T'.}
\end{deluxetable}

%% file: tab6.tex
\begin{deluxetable}{llcccccccllr}
\tablewidth{0pt}
\tabletypesize{\tiny}
\tablecaption{Optical Matches to UV Variable Sources in XMM/LSS (D1)\label{xmm_tab2}}
\tablehead{
\colhead{No.} & \colhead{\textsl{GALEX} ID} & \colhead{$u$} & \colhead{$g$} & \colhead{$r$} & \colhead{$i$} & \colhead{$z$} & \colhead{$r_{1/2}$} & \colhead{sep} & \colhead{flag\tablenotemark{a}} & \colhead{class\tablenotemark{b}} & \colhead{z} \\
\colhead{} & \colhead{} & \colhead{} & \colhead{} & \colhead{} & \colhead{} & \colhead{} & \colhead{(arcsec)} & \colhead{(arcsec)} & \colhead{} & \colhead{} & \colhead{}
}
\startdata
D1-1 & J022409.7-044007 & 22.111$\pm$0.007 & 21.288$\pm$0.003 & 20.286$\pm$0.002
 & 19.665$\pm$0.001 & 19.298$\pm$0.002 &  1.043 & 0.6 & o    & GAL & \\
D1-2 & J022409.7-043405 & 21.982$\pm$0.003 & 21.703$\pm$0.002 & 21.359$\pm$0.002
 & 21.298$\pm$0.002 & 21.348$\pm$0.008 &  0.461 & 1.0 &  x   & QSO & \\
D1-3 & J022424.2-043230 & 19.550$\pm$0.000 & 19.192$\pm$0.000 & 19.171$\pm$0.000
 & 18.923$\pm$0.000 & 18.972$\pm$0.001 &  0.452 & 1.0 &  x   & QSO & \\
D1-4 & J022434.7-045350 & 18.111$\pm$0.000 & 17.102$\pm$0.000 & 16.703$\pm$0.000
 & 16.612$\pm$0.000 & 16.420$\pm$0.000 &  0.500 & 2.2 &      & STAR & \\
D1-5 & J022449.2-041802 & 22.476$\pm$0.012 & 21.750$\pm$0.005 & 21.032$\pm$0.003
 & 20.419$\pm$0.002 & 20.132$\pm$0.006 &  0.703 & 0.2 &  x   & AGN & \\
D1-6 & J022449.9-043025 & 20.613$\pm$0.001 & 20.319$\pm$0.001 & 20.002$\pm$0.001
 & 19.982$\pm$0.001 & 19.850$\pm$0.002 &  0.472 & 0.2 &  x   & QSO & \\
D1-7 & J022451.2-040527 & 20.754$\pm$0.001 & 20.521$\pm$0.001 & 20.491$\pm$0.001
 & 20.276$\pm$0.001 & 20.278$\pm$0.004 &  0.458 & 0.6 &      & QSO & \\
D1-8 & J022503.2-040539 & 21.837$\pm$0.004 & 21.090$\pm$0.002 & 20.980$\pm$0.002
 & 20.791$\pm$0.002 & 20.458$\pm$0.005 &  0.564 & 0.9 &  x   & QSO & \\
D1-9 & J022517.0-043258 & 23.243$\pm$0.015 & 22.106$\pm$0.005 & 20.950$\pm$0.002
 & 20.369$\pm$0.001 & 20.064$\pm$0.004 &  0.636 & 0.1 & o    & GAL & \\
D1-10 & J022518.1-043155 & 24.450$\pm$0.039 & 24.272$\pm$0.032 & 23.360$\pm$
0.016 & 22.917$\pm$0.013 & 23.059$\pm$0.051 &  0.712 & 0.6 & o    & GAL & \\
D1-11 & J022526.5-040305 & 22.736$\pm$0.007 & 22.636$\pm$0.006 & 22.114$\pm$
0.004 & 21.947$\pm$0.004 & 21.362$\pm$0.009 &  0.528 & 0.2 & o    & QSO & \\
D1-12 & J022527.5-035954 & 21.528$\pm$0.003 & 21.254$\pm$0.002 & 21.225$\pm$
0.003 & 20.951$\pm$0.002 & 20.852$\pm$0.007 &  0.439 & 0.3 &      & QSO & \\
D1-13 & J022539.4-042228 & 21.955$\pm$0.004 & 21.457$\pm$0.002 & 21.269$\pm$
0.002 & 20.953$\pm$0.002 & 20.567$\pm$0.005 &  0.508 & 0.7 &      & QSO & \\
D1-14 & J022544.4-040102 & 22.042$\pm$0.005 & 21.637$\pm$0.004 & 21.396$\pm$
0.003 & 20.973$\pm$0.003 & 20.700$\pm$0.007 &  0.623 & 0.4 &      & GAL & \\
D1-15 & J022552.2-042249 & 21.417$\pm$0.002 & 21.056$\pm$0.001 & 20.982$\pm$
0.002 & 21.015$\pm$0.002 & 20.697$\pm$0.005 &  0.437 & 0.5 &  x   & QSO & \\
D1-16 & J022555.4-043918 & 21.069$\pm$0.002 & 20.823$\pm$0.001 & 20.568$\pm$
0.001 & 20.517$\pm$0.001 & 20.273$\pm$0.003 &  0.437 & 0.3 &  x   & QSO & \\
D1-17 & J022556.9-045853 & 21.484$\pm$0.003 & 21.109$\pm$0.002 & 20.790$\pm$
0.002 & 20.751$\pm$0.002 & 20.591$\pm$0.006 &  0.515 & 0.7 &      & QSO & \\
D1-18 & J022558.1-045720 & 20.962$\pm$0.002 & 20.677$\pm$0.001 & 20.386$\pm$
0.001 & 20.378$\pm$0.001 & 20.238$\pm$0.004 &  0.497 & 0.6 &      & QSO & \\
D1-19 & J022606.7-045723 & 21.004$\pm$0.002 & 20.709$\pm$0.001 & 20.629$\pm$
0.001 & 20.264$\pm$0.001 & 20.026$\pm$0.003 &  0.560 & 0.2 & ox   & QSO & \\
D1-20 & J022639.8-042004 & 21.525$\pm$0.002 & 21.640$\pm$0.003 & 21.430$\pm$
0.003 & 21.130$\pm$0.002 & 21.111$\pm$0.008 &  0.419 & 0.2 &  x   & QSO & \\
D1-21 & J022647.8-043135 & 22.720$\pm$0.007 & 22.518$\pm$0.006 & 22.052$\pm$
0.004 & 21.851$\pm$0.004 & 21.410$\pm$0.010 &  0.502 & 0.7 &      & QSO & \\
D1-22 & J022701.5-040913 & 21.389$\pm$0.003 & 21.021$\pm$0.002 & 20.701$\pm$
0.002 & 20.227$\pm$0.001 & 19.964$\pm$0.003 &  0.731 & 0.5 & ox   & AGN & \\
D1-23 & J022716.1-044539 & 20.939$\pm$0.002 & 20.320$\pm$0.001 & 20.309$\pm$
0.001 & 20.152$\pm$0.001 & 19.856$\pm$0.003 &  0.543 & 0.5 &  x   & QSO & \\
\enddata
\tablenotetext{a}{Sources that are optically variable are flagged with an 'o', 
sources detected as a hard X-ray point source are flagged with an 'x', sources
with an optical AGN spectrum are flagged with an 's', 
and sources with a UV quasar spectrum are flagged with a 'g'.}
\tablenotetext{b}{Optically unresolved sources ($r_{1/2} < 0\farcs6$) 
with $g-r < 0.6$ and $u-g < 1$ are classified as quasars (QSO).  Optically unresolved
sources with $u-g > 1.75$ or $g-r < 0.6$ and $u-g > 1$ are classified as stars (STAR).
Optically resolved sources ($r_{1/2} > 0\farcs6$) are classified as galaxies (GAL),
and optically resolved sources with a hard X-ray detection or an active galaxy spectrum
are classified as active galactic nuclei (AGN).}
\end{deluxetable}

%% file: tab7.tex
\begin{deluxetable}{llcccccccllr}
\tablewidth{0pt}
\tabletypesize{\tiny}
\tablecaption{Optical Matches to UV Variable Sources in COSMOS (D2)\label{cosmos_03_tab2}}
\tablehead{
\colhead{No.} & \colhead{\textsl{GALEX} ID} & \colhead{$u$} & \colhead{$g$} & \colhead{$r$} & \colhead{$i$} & \colhead{$z$} & \colhead{$r_{1/2}$} & \colhead{sep} & \colhead{flag\tablenotemark{a}} & \colhead{class\tablenotemark{b}} & \colhead{z} \\
\colhead{} & \colhead{} & \colhead{} & \colhead{} & \colhead{} & \colhead{} & \colhead{} & \colhead{(arcsec)} & \colhead{(arcsec)} & \colhead{} & \colhead{} & \colhead{}
}
\startdata
D2-1 & J095829.3+021542 & 21.665$\pm$0.011 & 21.297$\pm$0.005 & 21.000$\pm$0.005
 & 20.617$\pm$0.004 & 20.284$\pm$0.008 &  0.586 & 0.7 &      & QSO &  \\
D2-2 & J095848.9+023441 & 20.421$\pm$0.002 & 20.849$\pm$0.002 & 20.606$\pm$0.002
 & 20.344$\pm$0.002 & 20.456$\pm$0.005 &  0.498 & 0.5 &  xs  & QSO & 1.551\\
D2-3 & J095858.6+021459 & 19.486$\pm$0.006 & 18.054$\pm$0.001 & 17.253$\pm$0.001
 & 16.779$\pm$0.000 & 16.528$\pm$0.001 &  2.052 & 0.4 &  xsg & AGN & 0.132\\
D2-4 & J095921.5+024029 & 19.504$\pm$0.003 & 18.614$\pm$0.001 & 17.794$\pm$0.001
 & 17.308$\pm$0.000 & 17.002$\pm$0.001 &  1.949 & 2.0 &   sg & AGN & 0.260\\
D2-5 & J095947.0+022209 & 21.304$\pm$0.004 & 20.749$\pm$0.002 & 20.697$\pm$0.002
 & 20.763$\pm$0.002 & 20.554$\pm$0.006 &  0.472 & 0.7 &  xs  & QSO & 0.909\\
D2-6 & J095954.8+021706 & 17.345$\pm$0.001 & 15.458$\pm$0.000 & 14.634$\pm$0.000
 & 14.320$\pm$0.000 & 13.625$\pm$0.000 &  0.500 & 1.3 &      & STAR &  \\
D2-7 & J095958.5+021530 & 21.719$\pm$0.013 & 21.501$\pm$0.004 & 21.043$\pm$0.003
 & 20.145$\pm$0.002 & 19.794$\pm$0.003 &  0.731 & 0.8 &  x & AGN &  \\
D2-8 & J095958.9+020023 & 23.167$\pm$0.020 & 22.621$\pm$0.007 & 22.146$\pm$0.006
 & 21.831$\pm$0.005 & 21.459$\pm$0.010 &  0.515 & 1.1 & o    & QSO &  \\
D2-9 & J100002.0+024216 & 22.575$\pm$0.018 & 21.880$\pm$0.006 & 21.410$\pm$0.004
 & 20.771$\pm$0.003 & 20.464$\pm$0.007 &  0.655 & 0.4 &      & GAL &  \\
D2-10 & J100012.9+023522 & 19.454$\pm$0.001 & 18.941$\pm$0.000 & 18.868$\pm$
0.000 & 18.605$\pm$0.000 & 18.473$\pm$0.001 &  0.500 & 0.4 &  xsg & QSO & 0.699
\\
D2-11 & J100014.9+022717 & 21.882$\pm$0.011 & 21.114$\pm$0.004 & 20.964$\pm$
0.005 & 20.638$\pm$0.004 & 20.356$\pm$0.007 &  0.658 & 0.3 &      & GAL &  \\
D2-12 & J100017.6+020011 & 22.148$\pm$0.010 & 21.993$\pm$0.005 & 21.095$\pm$
0.003 & 20.664$\pm$0.002 & 20.299$\pm$0.005 &  0.569 & 1.2 &      &   &  \\
D2-13 & J100024.7+023149 & 18.915$\pm$0.000 & 18.814$\pm$0.000 & 18.614$\pm$
0.000 & 18.663$\pm$0.000 & 18.748$\pm$0.001 &  0.456 & 0.5 &  xsg & QSO & 1.318
\\
D2-14 & J100029.7+022129 & 21.754$\pm$0.007 & 21.023$\pm$0.002 & 20.775$\pm$
0.002 & 20.217$\pm$0.002 & 19.923$\pm$0.004 &  0.729 & 0.2 & o    & GAL &  \\
D2-15 & J100035.0+020235 & 21.079$\pm$0.004 & 20.862$\pm$0.002 & 20.549$\pm$
0.002 & 20.478$\pm$0.002 & 20.402$\pm$0.005 &  0.476 & 0.8 &  xs  & QSO & 1.177
\\
D2-16 & J100046.8+020404 & 20.701$\pm$0.004 & 20.386$\pm$0.002 & 20.043$\pm$
0.002 & 19.549$\pm$0.001 & 19.324$\pm$0.003 &  0.831 & 0.7 & oxs  & AGN & 0.554
\\
D2-17 & J100055.4+023441 & 20.802$\pm$0.007 & 20.558$\pm$0.001 & 20.187$\pm$
0.001 & 20.143$\pm$0.001 & 20.135$\pm$0.004 &  0.458 & 0.5 &  xs  & QSO & 1.403
\\
D2-18 & J100104.2+023349 & 25.030$\pm$0.273 & 24.565$\pm$0.041 & 24.468$\pm$
0.047 & 24.196$\pm$0.043 & 23.765$\pm$0.079 &  0.497 & 1.7 &      & QSO &  \\
D2-19 & J100104.3+023402 & 22.452$\pm$0.046 & 21.838$\pm$0.006 & 21.150$\pm$
0.004 & 20.882$\pm$0.004 & 20.739$\pm$0.009 &  0.900 & 0.8 &      & GAL &  \\
D2-20 & J100104.8+023329 & 25.831$\pm$0.566 & 25.902$\pm$0.141 & 25.842$\pm$
0.170 & 25.798$\pm$0.189 & 24.989$\pm$0.246 &  0.374 & 2.9 &      & QSO &  \\
D2-21 & J100105.4+023323 & 22.652$\pm$0.051 & 21.383$\pm$0.004 & 20.512$\pm$
0.002 & 20.082$\pm$0.002 & 19.800$\pm$0.004 &  0.683 & 1.0 &      & ART &  \\
D2-22 & J100116.3+023607 & 21.666$\pm$0.005 & 20.920$\pm$0.002 & 20.785$\pm$
0.002 & 20.608$\pm$0.002 & 20.308$\pm$0.004 &  0.582 & 0.4 &      & QSO &  \\
D2-23 & J100124.0+021445 & 22.887$\pm$0.014 & 22.241$\pm$0.005 & 22.120$\pm$
0.005 & 21.868$\pm$0.005 & 21.524$\pm$0.011 &  0.484 & 0.3 &  xs  & QSO & 0.894
\\
\enddata
\tablenotetext{a}{Sources that are optically variable are flagged with an 'o', 
sources detected as a hard X-ray point source are flagged with an 'x', sources
with an optical AGN spectrum are flagged with an 's', 
and sources with a UV quasar spectrum are flagged with a 'g'.}
\tablenotetext{b}{Optically unresolved sources ($r_{1/2} < 0\farcs6$) 
with $g-r < 0.6$ and $u-g < 1$ are classified as quasars (QSO).  Optically unresolved
sources with $u-g > 1.75$ or $g-r < 0.6$ and $u-g > 1$ are classified as stars (STAR).
Optically resolved sources ($r_{1/2} > 0\farcs6$) are classified as galaxies (GAL),
and optically resolved sources with a hard X-ray detection or an active galaxy spectrum are classified as active galactic nuclei (AGN).}
\end{deluxetable}

%% file: tab8.tex
\begin{deluxetable}{llcccccccllr}
\tablewidth{0pt}
\tabletypesize{\tiny}
\tablecaption{Optical Matches to UV Variable Sources in GROTH (D3)\label{groth_tab2}}
\tablehead{
\colhead{No.} & \colhead{\textsl{GALEX} ID} & \colhead{$u$} & \colhead{$g$} & \colhead{$r$} & \colhead{$i$} & \colhead{$z$} & \colhead{$r_{1/2}$} & \colhead{sep} & \colhead{flag\tablenotemark{a}} & \colhead{class\tablenotemark{b}} & \colhead{z} \\
\colhead{} & \colhead{} & \colhead{} & \colhead{} & \colhead{} & \colhead{} & \colhead{} & \colhead{(arcsec)} & \colhead{(arcsec)} & \colhead{} & \colhead{} & \colhead{}
}
\startdata
D3-1 & J141714.4+525130 & 26.621$\pm$0.332 & 25.568$\pm$0.071 & 25.243$\pm$0.060
 & 24.982$\pm$0.060 & 26.445$\pm$0.787 &  0.439 & 1.7 &      & STAR &  \\
D3-2 & J141715.2+530303 & 20.979$\pm$0.002 & 20.915$\pm$0.001 & 20.865$\pm$0.001
 & 20.784$\pm$0.002 & 20.550$\pm$0.004 &  0.472 & 0.6 & o    & QSO &  \\
D3-3 & J141721.2+530916 & 22.108$\pm$0.013 & 21.333$\pm$0.004 & 20.061$\pm$0.001
 & 19.503$\pm$0.001 & 19.162$\pm$0.002 &  0.945 & 0.4 & o    & GAL &  \\
D3-4 & J141724.6+523025 & 20.033$\pm$0.001 & 19.686$\pm$0.001 & 19.586$\pm$0.001
 & 19.324$\pm$0.001 & 19.124$\pm$0.001 &  0.500 & 0.7 &  x g & QSO & 0.482\\
D3-5 & J141734.9+522811 & 21.988$\pm$0.006 & 21.951$\pm$0.003 & 21.410$\pm$0.002
 & 21.338$\pm$0.003 & 20.841$\pm$0.006 &  0.508 & 0.8 &  x   & QSO &  \\
D3-6 & J141801.5+525201 & 20.261$\pm$0.001 & 20.162$\pm$0.001 & 19.991$\pm$0.001
 & 19.963$\pm$0.001 & 19.849$\pm$0.003 &  0.474 & 0.3 &    g & QSO & 1.159\\
D3-7 & J141830.3+522213 & 21.520$\pm$0.004 & 20.740$\pm$0.001 & 20.713$\pm$0.001
 & 20.546$\pm$0.001 & 20.243$\pm$0.004 &  0.523 & 0.6 &  x g & QSO & 0.825\\
D3-8 & J141833.5+530733 & 22.957$\pm$0.016 & 22.365$\pm$0.006 & 21.726$\pm$0.004
 & 20.969$\pm$0.002 & 20.666$\pm$0.006 &  0.642 & 0.3 & o    & GAL &  \\
D3-9 & J141834.9+524205 & 25.825$\pm$0.196 & 25.581$\pm$0.075 & 24.649$\pm$0.035
 & 24.319$\pm$0.033 & 23.860$\pm$0.079 &  0.458 & 1.6 &      &   &  \\
D3-10 & J141903.9+530856 & 22.533$\pm$0.009 & 21.925$\pm$0.003 & 21.692$\pm$
0.003 & 21.411$\pm$0.003 & 21.484$\pm$0.011 &  0.441 & 0.4 &      & QSO &  \\
D3-11 & J141905.2+522528 & 19.560$\pm$0.001 & 19.539$\pm$0.001 & 19.503$\pm$
0.001 & 19.322$\pm$0.001 & 19.313$\pm$0.002 &  0.437 & 0.7 &   sg & QSO & 1.606
\\
D3-12 & J141918.1+524158 & 20.426$\pm$0.001 & 20.322$\pm$0.001 & 20.297$\pm$
0.001 & 20.149$\pm$0.001 & 19.930$\pm$0.003 &  0.430 & 0.3 &    g & QSO & 2.034
\\
D3-13 & J141929.8+525206 & 22.741$\pm$0.025 & 21.363$\pm$0.004 & 20.108$\pm$
0.002 & 19.634$\pm$0.001 & 19.349$\pm$0.004 &  0.625 & 0.4 & o    & GAL &  \\
D3-14 & J141936.1+521359 & 23.179$\pm$0.021 & 22.141$\pm$0.005 & 21.374$\pm$
0.003 & 20.985$\pm$0.002 & 20.762$\pm$0.007 &  0.688 & 1.5 &      & ERR &  \\
D3-15 & J141946.4+525943 & 17.619$\pm$0.000 & 15.695$\pm$0.000 & 14.828$\pm$
0.000 & 14.138$\pm$0.000 & 13.443$\pm$0.000 &  0.500 & 0.5 &      & STAR &  \\
D3-16 & J141951.5+524205 & 22.711$\pm$0.016 & 22.066$\pm$0.006 & 21.404$\pm$
0.003 & 20.678$\pm$0.002 & 20.405$\pm$0.006 &  0.766 & 0.6 & o    & GAL &  \\
D3-17 & J141959.8+524245 & 22.253$\pm$0.008 & 21.403$\pm$0.002 & 20.649$\pm$
0.001 & 20.255$\pm$0.001 & 20.039$\pm$0.003 &  0.508 & 0.2 & o    &   &  \\
D3-18 & J142039.5+521929 & 19.519$\pm$0.001 & 19.323$\pm$0.000 & 18.985$\pm$
0.000 & 18.665$\pm$0.000 & 18.587$\pm$0.001 &  0.437 & 0.6 &   sg & QSO & 1.590
\\
D3-19 & J142048.2+530816 & 23.139$\pm$0.013 & 23.791$\pm$0.014 & 23.672$\pm$
0.015 & 23.489$\pm$0.015 & 22.899$\pm$0.030 &  0.450 & 0.3 &      & QSO &  \\
D3-20 & J142056.9+524830 & 20.653$\pm$0.002 & 20.548$\pm$0.001 & 20.187$\pm$
0.001 & 20.127$\pm$0.001 & 20.074$\pm$0.004 &  0.458 & 0.3 &    g & QSO & 1.199
\\
D3-21 & J142107.7+530319 & 22.383$\pm$0.009 & 21.871$\pm$0.004 & 21.529$\pm$
0.003 & 20.958$\pm$0.002 & 20.639$\pm$0.005 &  0.629 & 0.9 & o    & GAL &  \\
D3-22 & J142112.3+524147 & 21.111$\pm$0.003 & 20.446$\pm$0.001 & 20.369$\pm$
0.001 & 20.298$\pm$0.001 & 20.085$\pm$0.003 &  0.506 & 0.1 &    g & QSO & 0.847
\\
D3-23 & J142113.9+521747 & 19.934$\pm$0.001 & 19.715$\pm$0.001 & 19.703$\pm$
0.001 & 19.565$\pm$0.001 & 19.451$\pm$0.002 &  0.459 & 0.6 &      & QSO &  \\
D3-24 & J142115.7+523156 & 21.674$\pm$0.004 & 21.301$\pm$0.002 & 21.062$\pm$
0.002 & 20.933$\pm$0.002 & 20.913$\pm$0.007 &  0.428 & 0.8 &      & QSO &  \\
D3-25 & J142125.0+521800 & 22.971$\pm$0.013 & 22.523$\pm$0.005 & 22.299$\pm$
0.005 & 21.989$\pm$0.005 & 21.483$\pm$0.010 &  0.497 & 0.5 &      & QSO &  \\
D3-26 & J142133.9+530245 & 22.141$\pm$0.006 & 21.702$\pm$0.003 & 21.556$\pm$
0.003 & 21.350$\pm$0.003 & 21.040$\pm$0.007 &  0.487 & 0.2 &      & QSO &  \\
D3-27 & J142134.2+523429 & 21.194$\pm$0.003 & 21.007$\pm$0.001 & 20.688$\pm$
0.001 & 20.611$\pm$0.002 & 20.435$\pm$0.004 &  0.465 & 0.5 &    g & QSO & 1.199
\\
D3-28 & J142135.9+523139 & 19.601$\pm$0.002 & 19.050$\pm$0.001 & 18.298$\pm$
0.001 & 17.859$\pm$0.000 & 17.624$\pm$0.001 &  1.326 & 0.6 & o sg & AGN & 1.249
\\
D3-29 & J142151.1+524951 & 20.786$\pm$0.005 & 19.615$\pm$0.001 & 18.818$\pm$
0.001 & 18.355$\pm$0.001 & 18.111$\pm$0.001 &  1.196 & 0.3 & o    & GAL &  \\
D3-30 & J142206.8+524957 & 22.740$\pm$0.009 & 22.076$\pm$0.003 & 22.139$\pm$
0.004 & 21.690$\pm$0.003 & 21.853$\pm$0.014 &  0.439 & 0.8 &      & QSO &  \\
D3-31 & J142209.1+530559 & 19.907$\pm$0.001 & 19.515$\pm$0.001 & 19.429$\pm$
0.001 & 19.207$\pm$0.001 & 18.902$\pm$0.001 &  0.597 & 0.1 & o  g & QSO & 0.760
\\
D3-32 & J142243.5+530520 & 20.401$\pm$0.005 & 20.407$\pm$0.003 & 20.122$\pm$
0.003 & 20.053$\pm$0.003 & 19.938$\pm$0.011 &  0.621 & 0.3 &      & GAL &  \\
\enddata
\tablenotetext{a}{Sources that are optically variable are flagged with an 'o', 
sources detected as a hard X-ray point source are flagged with an 'x', sources
with an optical AGN spectrum are flagged with an 's', 
and sources with a UV quasar spectrum are flagged with a 'g'.}
\tablenotetext{b}{Optically unresolved sources ($r_{1/2} < 0\farcs6$) 
with $g-r < 0.6$ and $u-g < 1$ are classified as quasars (QSO).  Optically unresolved
sources with $u-g > 1.75$ or $g-r < 0.6$ and $u-g > 1$ are classified as stars (STAR).
Optically resolved sources ($r_{1/2} > 0\farcs6$) are classified as galaxies (GAL),
and optically resolved sources with a hard X-ray detection or an active galaxy spectrum
are classified as active galactic nuclei (AGN).}
\end{deluxetable}

%% file: tab9.tex
\begin{deluxetable}{llcccccccllr}
\tablewidth{0pt}
\tabletypesize{\tiny}
\tablecaption{Optical Matches to UV Variable Sources in CFHTLS D4\label{d4_tab2}}
\tablehead{
\colhead{No.} & \colhead{\textsl{GALEX} ID} & \colhead{$u$} & \colhead{$g$} & \colhead{$r$} & \colhead{$i$} & \colhead{$z$} & \colhead{$r_{1/2}$} & \colhead{sep} & \colhead{flag\tablenotemark{a}} & \colhead{class\tablenotemark{b}} & \colhead{z} \\
\colhead{} & \colhead{} & \colhead{} & \colhead{} & \colhead{} & \colhead{} & \colhead{} & \colhead{(arcsec)} & \colhead{(arcsec)} & \colhead{} & \colhead{} & \colhead{}
}
\startdata
D4-1 & J221336.1-180531 & 19.459$\pm$0.000 & 19.169$\pm$0.000 & 19.208$\pm$0.000
 & 18.977$\pm$0.000 & 19.001$\pm$0.001 &  0.484 & 0.3 & o    & QSO & \\
D4-2 & J221409.7-175652 & 20.046$\pm$0.001 & 19.894$\pm$0.001 & 19.639$\pm$0.001
 & 19.541$\pm$0.001 & 19.538$\pm$0.002 &  0.465 & 0.2 &      & QSO & \\
D4-3 & J221413.0-172909 & 20.730$\pm$0.002 & 20.584$\pm$0.002 & 20.251$\pm$0.002
 & 20.163$\pm$0.002 & 20.072$\pm$0.004 &  0.487 & 1.2 & o    & QSO & \\
D4-4 & J221429.5-173007 & 18.455$\pm$0.000 & 18.126$\pm$0.000 & 18.104$\pm$0.000
 & 18.126$\pm$0.000 & 17.960$\pm$0.001 &  0.459 & 0.2 &      & QSO & \\
D4-5 & J221432.7-180158 & 20.253$\pm$0.001 & 19.812$\pm$0.001 & 19.684$\pm$0.001
 & 19.638$\pm$0.001 & 19.382$\pm$0.002 &  0.484 & 0.1 &      & QSO & \\
D4-6 & J221433.8-173007 & 21.686$\pm$0.006 & 20.844$\pm$0.002 & 19.962$\pm$0.001
 & 19.524$\pm$0.001 & 19.207$\pm$0.003 &  0.831 & 0.0 & o    & GAL & \\
D4-7 & J221508.9-175602 & 25.067$\pm$0.075 & 24.499$\pm$0.039 & 24.584$\pm$0.048
 & 23.842$\pm$0.030 & 23.438$\pm$0.069 &  0.746 & 0.8 & o    & GAL & \\
D4-8 & J221639.6-174427 & 20.916$\pm$0.002 & 20.396$\pm$0.001 & 20.385$\pm$0.001
 & 19.995$\pm$0.001 & 19.743$\pm$0.003 &  0.532 & 0.3 &      & QSO & \\
D4-9 & J221647.8-175454 & 20.765$\pm$0.002 & 20.358$\pm$0.001 & 20.138$\pm$0.001
 & 19.673$\pm$0.001 & 19.257$\pm$0.002 &  0.595 & 0.4 &      & QSO & \\
\enddata
\tablenotetext{a}{Sources that are optically variable are flagged with an 'o', 
sources detected as a hard X-ray point source are flagged with an 'x', sources
with an optical AGN spectrum are flagged with an 's', 
and sources with a UV quasar spectrum are flagged with a 'g'.}
\tablenotetext{b}{Optically unresolved sources ($r_{1/2} < 0\farcs6$) 
with $g-r < 0.6$ and $u-g < 1$ are classified as quasars (QSO).  Optically unresolved
sources with $u-g > 1.75$ or $g-r < 0.6$ and $u-g > 1$ are classified as stars (STAR).
Optically resolved sources ($r_{1/2} > 0\farcs6$) are classified as galaxies (GAL),
and optically resolved sources with a hard X-ray detection or an active galaxy spectrum
are classified as active galactic nuclei (AGN).}
\end{deluxetable}

%% file: tab10.tex
\begin{deluxetable}{llclllll}
\tablewidth{0pt}
\tablecaption{Tidal Disruption Flare Candidates \label{tab_cand}}
\tablehead{
\colhead{No.} & \colhead{\textsl{GALEX} ID} & \colhead{flare} & \colhead{flag} & \colhead{temp\tablenotemark{a}} & \colhead{photo-$z$} & \colhead{spec-$z$} & \colhead{spec type\tablenotemark{b}}\\
}
\startdata
D1-1 & J022409.7-044007 & F & o   & em & 0.52 - 0.60 & 0.436 & Sy2 \\
D1-9 & J022517.0-043258 & T & o   & Sb & 0.28 - 0.36 & 0.326 & early \\
D1-10 & J022518.1-043155 & F & o   & em  & 0.48 - 0.60 & \nodata & \nodata \\
D1-14 & J022544.4-040102 & F &     & em  & 1.36 - 1.48 & \nodata & \nodata \\
D2-9 & J100002.0+024216 & F &     & em  & 0.80 - 0.88 & 0.579 & Sy2\\
D2-11 & J100014.9+022717 & F &     & em & 0.00 - 0.08 & 0.727 & Sy1 \\
D3-3 & J141721.2+530916 & T & o   & Sa  & 0.48 - 0.60 & 0.355 & Sy2 \\
D3-8 & J141833.5+530733 & T & o   & em  & 0.72 - 0.80 & 0.681 & ? \\
D3-13 & J141929.8+525206 & T & o   & E/S0  & 0.28 - 0.36 & 0.3698 & early\\
D3-21 & J142107.7+530319 & F & o   & em & 0.96 - 1.04 & 0.735 & Sy1\\
D4-7 & J221508.9-175602 & F & o   & em  & 0.96 - 1.12 & 0.189 & SN IIn\\
\enddata
\tablenotetext{a}{Galaxy template from photo-z fitting routine.  Emission line galaxies (either star forming or Seyfert galaxies) are labeled as `em'.}
\tablenotetext{b}{Spectra with no emission lines are classified as early type galaxies (early).  Spectra with diagnostic narrow-line ratios of ([O~III] $\lambda 5007$)/(H$\beta) > 3$ are classified as Seyfert 2 galaxies (Sy2) and spectra that show broad H$\beta$ emission are classified as Seyfert 1 galaxies (Sy1).}
\end{deluxetable}

%% file: tab11.tex
\begin{deluxetable}{llllllrllrr}
\tablewidth{0pt}
\tablecaption{Log of Optical Spectrsocopic Observations \label{tab:spec}}
\tablehead{
\colhead{No.} & \colhead{\textsl{GALEX} ID} & \colhead{UT Date} & \colhead{Telescope} 
& \colhead{Instr.} & \colhead{slit} & \colhead{$\Delta \lambda$} & \colhead{t$_{\rm exp}$} & \colhead{z} & \colhead{[O~III]/H$\beta$} & \colhead{log(L([O~III]))} \\
\colhead{} & \colhead{} & \colhead{} & \colhead{} & \colhead{} & \colhead{($\arcsec$)} & \colhead{(\AA)} & \colhead{(s)} & \colhead{} & \colhead{} & \colhead{(ergs s$^{-1}$)}
}
\startdata
D1-1 & J022409.1-044007 & 2007-09-16 & Keck & LRIS & 1.0 & 5.7 & 1800 & 0.436 & 4.0 & 41.3\\
D1-9 & J022517.0-043258 & 2006-09-26 & VLT & FORS1 & 0.7 & 5.0 & 2250 & 0.326 & \nodata & $< $37.8\\
D2-9 & J100002.0+024216 & 2007-05-18 & Keck & LRIS & 0.7 & 4.8 & 2700 & 0.579 & 3.6 & 41.1\\
D2-11 & J100014.9+022717 & 2007-05-18 & Keck & LRIS & 0.7 & 4.8 & 1800 & 0.727 & 2.0 & 41.4 \\
D3-3 & J141721.2+530916 & 2005-12-28 & MDM 2.4m & CCDS & 2.0 & 15.0 & 2700 & 0.355 & 9.0 & 40.4\\
D3-8 & J141833.5+530733 & 2007-04-19 & Keck & LRIS & 1.0 & 5.7 & 2700  & 0.681 & 2.7 & 40.4\\
D3-13 & J141929.8+525206 & 2007-04-19 & Keck & LRIS & 1.0 & 5.7 & 2400  & 0.3698 & \nodata & $<$ 39.6 \\
D3-21 & J132107.7+530319 & 2007-04-19 & Keck & LRIS & 1.0 & 5.7 & 3600 & 0.735 & $>$ 2.0 & 40.6 \\
\enddata
\end{deluxetable}